\documentclass[a4paper]{revtex4}
\usepackage{inputenc}
\usepackage[english]{babel}

\usepackage{array,booktabs}% http://ctan.org/pkg/{array,booktabs}
\usepackage{array} 
\usepackage{lipsum}   
\usepackage{calc}
\RequirePackage{graphicx}
\usepackage{color}
\usepackage{amsmath}
\usepackage{float}
\usepackage{calc}
\usepackage{pdflscape}
\usepackage{color}
\usepackage{float}
\usepackage{subfigure}
\usepackage[font=small,labelfont=bf]{caption}
\usepackage{graphicx}
\usepackage{pdflscape}
\usepackage{color}
\usepackage{float}
\usepackage[font=small,labelfont=bf]{caption}
\usepackage{graphicx}
\usepackage{amsmath}
\usepackage[nodisplayskipstretch]{setspace}
\renewcommand{\arraystretch}{1.5}
\usepackage{setspace}
\usepackage{tabularx}

\usepackage{float}
\usepackage{color}
\usepackage{amsmath}
\usepackage{float}
\usepackage{calc}
\usepackage{pdflscape}
\usepackage{color}
\usepackage{float}
\usepackage{subfigure}
\usepackage[font=small,labelfont=bf]{caption}
\usepackage{graphicx}
\begin{document}{\setlength\abovedisplayskip{4pt}}

\title{Non$-$zero $ \theta_{13} $ and $ \delta_{CP} $ phase with $ A_{4} $ Flavor Symmetry and Deviations to Tri$-$Bi$-$Maximal mixing via $ Z_{2} \times  Z_{2}$ invariant perturbations in the Neutrino sector.}

\author{Gayatri Ghosh}
\email{gayatrighdh@gmail.com}
\affiliation{Department of Physics, Gauhati University, Jalukbari, Assam-781015, India}
\affiliation{Department of Physics, Pandit Deendayal Upadhayay Mahavidyalaya, Karimganj, Assam-788720, India}

%\ead{{\color{blue}kalpana@gauhati.ac.in}}
%\ead{{\color{blue}gayatrighsh@gmail.com}}
%\ead{{e-mail: kalpana.bora@gmail.com}}
\begin{abstract}

In this work, a flavour theory of a neutrino mass model based on $ A_{4} $ 
symmetry is considered to explain the phenomenology of neutrino mixing. The spontaneous symmetry breaking of $ A_{4} $ symmetry in this model leads to tribimaximal mixing in the neutrino sector at a leading order. We consider the effect of $ Z_{2} \times  Z_{2}$ invariant perturbations in neutrino sector and find the allowed region of correction terms in the perturbation matrix that is consistent with 3$ \sigma $ ranges of the experimental values of the mixing angles. We study the entanglement of this formalism on the other phenomenological observables, such as $ \delta_{CP} $ phase, the neutrino oscillation probability $ P(\nu_{\mu}\rightarrow \nu_{e} )$, the effective Majorana mass $ |m_{ee} |$ and $ |m^{eff}_{\nu e} |$. A $ Z_{2} \times  Z_{2}$ invariant perturbations in this model is introduced in the neutrino sector which leads to testable predictions of $ \theta_{13} $ and CP violation. By changing the magnitudes of perturbations in neutrino sector, one can generate viable values of $ \delta_{CP} $ and neutrino oscillation parameters. Next we investigate the feasibility of charged lepton flavour violation in type-I seesaw models with leptonic flavour symmetries at high energy that leads to tribimaximal neutrino mixing. We consider an effective theory with an $A_{4} \times Z_{2} \times  Z_{2} $ symmetry, which after spontaneous symmetry breaking at high scale which is much higher than the electroweak scale leads to charged lepton flavour violation processes once the heavy Majorana neutrino mass degeneracy is lifted either by renormalization group effects or by a soft breaking of the $ A_{4} $ symmetry. In this context the implications for charged lepton flavour violation processes like $ \mu  \rightarrow e  \gamma $, $ \tau  \rightarrow e  \gamma $, $ \tau  \rightarrow \mu  \gamma $ are discussed.
\end{abstract}   
\maketitle
\section{Introduction}
\label{intro}
Ever since the discovery of neutrino oscillations, the aspects of lepton masses, mixings and flavour violation {\color{blue}\cite{GG}} have been an active topic of research and there have been a lot of updates on  the results from a long ongoing series of global fits to neutrino oscillation data {\color{blue}\cite{a,b,c,d}}. Neutrino flavor conversion was first detected in solar {\color{blue}\cite{1}} and atmospheric
neutrinos {\color{blue}\cite{2}}. This discovery led to the Nobel prize in Physics in 2015 {\color{blue}\cite{3,4}} and was confirmed by subsequent results from the KamLAND reactor experiment {\color{blue}\cite{5}} as well as long
baseline accelerator experiments. 
\par 
The neutrinos change their flavour as they propagate in space and this phenomenon is known as neutrino oscillation which occurs since the flavour gauge eigenstates
of neutrinos are mixture of mass eigenstates. The mixing is described by PMNS matrix which can be parameterized in terms of three neutrino mixing angles and CP violating
phases. The experimental disovery of neutrino oscillations constitutes not only neutrino mass squared differences, but the probability of nearly degenerate neutrino spectrum is also contemplated. Further neutrino oscillation has triggered the experimental and theoretical endeavour to understand the aspects of lepton masses, mixings and flavour violation in SUSY GUTs theories. The massive neutrinos are produced in their gauge 
eigenstates $\left(\nu_{\alpha}\right)$ which is related to their mass eigenstate $\left(\nu_{i}\right)$, where the gauge eigenstates take part in gauge interactions.
\begin{equation}
\mid \nu_{\alpha} > = \sum U_{\alpha_{i}}\mid \nu_{i} >
\end{equation} 
where, $\alpha = e, \mu, \tau$ , $\nu_{i}$ is the neutrino of distinct mass $ m_{i} $.
\par 
In the physics of the dynamics of neutrino mass generation in the leptonic sector, the flavour problem of particle physics, is one of the open challeneges that the field of high energy physics faces today.
\par 
Since the flavour mixing happens due to the mixing between mass and flavour eigenstates, neutrinos have nondegenerate mass. To put into effect this idea into a renormalisable field theory what so ever symmetry used in generating neutrino mass degeneracy must be broken. In this work $ A_{4} $ symmetry {\color{blue}\cite{39d,39,39a,39b,39c}} which is the  group of the even permutation of four objects or equivalently that of a tetrahedron used to  maintain this degeneracy is broken spontaneously to produce the spectrum of different charged lepton masses.
\par
Many inferences have been intended to guess the actual pattern of lepton mixings. Some of the phenomenological pattern of neutrino mixings incorporate for example, Tri-bimaximal (TBM) {\color{blue}\cite{39a,39b,39c}}, Trimaximal (TM1/TM2) and bi-large mixing patterns.
\par 
Over the past two decades, a lot of theoretical and experimental works have been going on, which aimed at grasping the structure of lepton mixing matrix {\color{blue}\cite{e}}. Solar and atmospheric angle as conferred by accelerator and reactor data indicated that the mixing in the lepton sector is very different from  quark mixings, given the large  values of $ \theta_{12} $ and $ \theta_{23} $. These observations were soon encrypted 
in the tribimaximal (TBM) mixing ansatz presented by Harrison, Perkins, and Scott {\color{blue}\cite{6}} and also {\color{blue}\cite{dhrm}} described by.

\begin{equation}
U_{PMNS} \simeq \begin{bmatrix}
\frac{2}{\sqrt{6}} & \frac{1}{\sqrt{3}} & 0\\
-\frac{1}{\sqrt{6}} & \frac{1}{\sqrt{3}}& -\frac{1}{\sqrt{2}}\\
-\frac{1}{\sqrt{6}} & -\frac{1}{\sqrt{3}} & \frac{1}{\sqrt{2}}\\
\end{bmatrix}
= U_{TBM}
\end{equation}
where, $ Sin\hspace{0.1cm}\theta_{13}=0 $. In this educated guess, mixing angles have $ Sin \hspace{0.1cm}\theta_{12} = \frac{1}{3}$, $ \theta_{23} = \frac{\pi}{4}$ and $ Sin ^{2}\hspace{0.1cm} \theta_{23}=\frac{1}{2}$ whose perspective is good bearing in mind the latest neutrino oscillation global fit. Since the TBM  ansatz was first proposed so it became a touchstone convention for inspiring the pattern of lepton masses and mixings. Unfortunately, it envisages $ Sin\hspace{0.1cm}\theta_{13}=0 $ and hence zero leptonic CP violation phase in neutrino oscillation. Infact, data from reactors have stipulated that such sterling TBM ansatz can not be the correct description of nature, since the reactor mixing angle $\theta_{13}$ has been confirmed to be non-zero to a very high significant content {\color{blue}\cite{7,8}}. Further, till now it is becoming increasingly apparent that there has been compelling evidence for CP violation in neutrino oscillations, allocating further hint that alteration or change of TBM mixing ansatz is vital.
\par 
Neutrino oscillation experiments are a probe to measure neutrino mixing and mass spectrum since the oscillation probability $P(\nu_{\mu}\rightarrow \nu_{e} )$ depends on mixing angles, Dirac CP Violation phase and the mass square differences $ m^{2}_{21}, m^{2}_{23} $. Results from earlier experiments stipulate that $\theta_{13} $ is very
small, almost zero and the lepton mixing matrix follows the TBM $($tri-bimaximal mixing$)$ ansatz. This ansatz tells $Sin\hspace{0.1cm}\theta_{13}=0$, $Sin^{2}\theta_{23}=\frac{1}{2}$, $Tan^{2}\theta_{12}=\frac{1}{2}$. One can conclude the neutrino mixing matrix as TBM type, with small deviations or corrections to it due to perturbation in the charged-lepton or neutrino sector. Current experimental observations of fairly large $ \theta_{13} $ {\color{blue}\cite{a,b,c,d}}, deviated neutrino mixing a little away from TBM ansatz, but close to the predictions of non-zero $ \theta_{13} $ and $ \delta_{CP} = \pm \frac{\pi}{2}$. One can correlate the CP violation in neutrino oscillation with the octant of the atmospheric mixing angle $ \theta_{23} $. In this paper, we would like to address a model based on $ A_{4} $ symmetry which gives non-zero $ \theta_{13} $, $ \delta_{CP} = \pm \frac{\pi}{2}$ and $ Sin^{2}\theta_{23} = 0.57$ via perturbations in the form of $ Z_{2}\times Z_{2} $ invariant symmetry in the neutrino sector at leading order. In order to take into account the deviations in mixing angles at a leading order  consistent with the experimental results, we add a perturbation in neutrino sector in the form of $ Z_{2}\times Z_{2} $ invariant symmetry including second order corrections in the PMNS matrix.
\par
The predictions of vanishing $ \theta_{13} $ by TBM is owing to its invariance under $ \mu-\tau $ exchange symmetry {\color{blue}\cite{GG}}. Small explicit breaking of $ \mu-\tau $ symmetry can generate large Dirac CP violating phase in neutrino oscillations {\color{blue}\cite{GC}}. Also some studies in the context of corrections to TBM mixing in   $ A_{4} $ symmetry is presented in {\color{blue}\cite{BK}}. All CP violations (both Dirac and Majorana types) emerge from a common origin in neutrino seesaw.  $ \mu-\tau $ symmetry breaking shares the common origin with all CP violations, since in the limit of $ \mu-\tau $ symmetry mixing angle $ \theta_{13} $ becomes non$-$zero and thus CP conservation takes place {\color{blue}\cite{Pcheng1}}. Studies on common origin of soft $\mu-\tau$ symmetry and CP breaking in neutrino seesaw, the origin of matter, baryon asymmetry, hidden flavor symmetry are vividly illustrated in {\color{blue}\cite{Pcheng1, Pcheng2}}. $ \mu-\tau $ symmetry and its breaking together with CP violation, correlation between $\theta_{13}$, $\theta_{23}$ and $\theta_{12}$, in connection to hidden flavor symmetry $($including $Z_2 \times Z_2$ $)$ are extensively studied in {\color{blue}\cite{Pcheng1, Pcheng2, Pcheng3, Pcheng4}}. Octahedral symmetric group $O_{h}$ is described as the flavor symmetry of neutrino-lepton sector. Here the residual symmetries are $ Z_{2}^{\mu-\tau} \bigotimes Z_{2}^{s}$ and $ Z_{l}^{4}$ and it prescribe the neutrinos and charged leptons, respectively {\color{blue}\cite{Pcheng3}}. Studies on the notion of constrained maximal CP violation (CMCPV) which predicts the features $\delta_{CP} =\frac{-\pi}{2}\hspace{0.1cm} \text{and}\hspace{0.1cm}\theta_{23}=\frac{\pi}{4}$ and their origin in the context of flavor symmetry is presented in {\color{blue}\cite{Pcheng4}}. With the discovery of non$-$zero value of the reactor mixing angle $ \theta_{13} $ by reactor experiments RENO {\color{blue}\cite{9}} and Daya Bay {\color{blue}\cite{10}} the texture of tri$-$bimaximal mixings can be generalised.
 \par 
A minimally asymmetric Yukawa texture based on the Frobenius group $T_{13}$ and SU$(5)$ GUT are presented in {\color{blue}\cite{pa,ka,si,ha}}, where the neutrino masses and mixing angles are predicted from TBM seesaw mixing, (up to a sign), the CP violating phase (1.32$\pi)$ and Jarlskog invariant are determined in agreement with current global fits with definite prediction for neutrinoless double beta experiments, and baryon asymmetry is explained from flavored leptogenesis. 
\par 
The essence of knowing the exact symmetries behind the observed pattern of neutrino oscillations is one of the challenging tasks in particle physics.
\par 
In this work we propose a $ A_{4} $ family symmetry {\color{blue}\cite{11}}  $ - $ the symmetry group of even permutations of 4 objects or equivalently that of a tetrahedron, which is used here to obtain neutrino mixing predictions within fundamental theories of neutrino mass. This $ A_{4} $ family symmetry was first introduced as a possible family symmetry for the quark sector {\color{blue}\cite{12}} and is now mostly used for
the lepton sector {\color{blue}\cite{13,14,15,16,17}}. During last two decades many neutrino oscillation experiments like KamLAND  {\color{blue}\cite{18}}, LBL+ATM+REAC {\color{blue}\cite{19}}, SOL, LBL+ATM, REAC, LBL, (LBL+REAC) and ATM {\color{blue}\cite{20}} are being performed and the oscillation parameters are being measured to a very good precision. On the light of discovery of non zero $ \theta_{13} $, the neutrino mass model dictating TBM mixing pattern needs necessary modifications.
\par 
The discrete family symmetry groups necessitates the need of special vacuum alignment condition to implement tribimaximal mixing pattern ansatz. Also one can generate deviations from TBM mixing pattern by adding symmetry breaking terms in the interactive Lagrangian of the specific discrete family symmetry group. This results in partial and complete symmetry breaking. Residual symmetries exist in neutrino and charged lepton sectors after such perturbations.
\par 
In this work we also carry out studies on lepton flavour violation decay $ \mu  \rightarrow e  \gamma $, $ \tau  \rightarrow \mu  \gamma $ and $ \tau  \rightarrow e  \gamma $ in $ G_{SM} \times A_{4}\times U(1)_{X}$ incorporating $ Z_{2} \times Z_{2}$ invariant perturbation in both charged lepton sector and neutrino sector as discussed below, and hence one can guess the sensitivity to test the observation of sleptons and sparticles at future run of LHC. These charged lepton flavour violation rates depend on the form of Dirac neutrino yukawa couplings as fixed by most favourable predicted value of Dirac CPV phase of this work and on the details of soft SUSY breaking parameters and Tan$\beta$. We have used the  Higgs mass as measured at LHC, non zero reactor mixing angle $ \theta_{13} $ for neutrinos, and latest present and future constraints on BR($ \mu  \rightarrow e  \gamma $) {\color{blue}\cite{14nov}}.
\par 
Persuaded by the prerequisite for departing from the simplest first$-$order form for the TBM ansatz, {\color{blue}Eq. (2)}, here we propose a generalized version of the TBM ansatz in which the new ansatz is realised in a model based on $ A_{4} $ group as suggested in {\color{blue}\cite{39c}} by breaking $ A_{4} $ symmetry spontaneously to $ Z_{2} $ in the neutrino sector, which correctly accounts for the non-zero value of
$ \theta_{13} $ and introduces CP violation. We then incorporate a real $ Z_{2}\times Z_{2} $ perturbations in the neutrino sector leading to feasible values of $ \theta_{13} $ and $ \delta_{CP} $. This results in predictions of neutrino oscillation parameters and leptonic CPV phase that will be tested at upcoming neutrino experiments. Appendix A summarizes the $ A_{4} $ algebra.
\section{The $ A_{4} $ model}

We take a type I SeeSaw model based on $ A_{4} $ symmetry {\color{blue}\cite{39c}}. Let us limit ourselves to only leptonic sector. The field consists of three left handed  $SU(2)_{L}$ gauge doublets, three right handed charged gauge singlets, three right handed neutrino gauge singlets. In addition there exists also four Higgs doublets $ \phi_{i} $ $\left( i=1,2,3\right)$ and $\phi_{0}$ and three scalar singlets. The above fields can be represented under various irreducible representations as:

\begin{table}[htb]
\renewcommand{\arraystretch}{1.5}
\begin{center}
\begin{tabular}{|c|c|c|c|c|}
\hline 
 Fields & $SU(2)_{L}$ & $U(1)_{Y}$& $ A_{4} $& Representation\\ 
\hline 
%%\br
 Left \hspace{0.1cm} Handed \hspace{0.1cm} Doublets & $\frac{1}{2}$ &$Y=-1$ &$\underline{3}$&$Y_{iL}$\\
\hline
Right\hspace{0.1cm} Handed \hspace{0.1cm}Charged\hspace{0.1cm}Lepton\hspace{0.1cm} Singlets & 0 &$Y=-2$ &$\underline{1}\oplus\underline{1^{'}}\oplus1^{''}$&$l_{iR}$\\
\hline
 Right\hspace{0.1cm} Handed \hspace{0.1cm}Neutrino\hspace{0.1cm} Singlets & 0 &$Y=0$ &$\underline{3}$&$\nu_{iR}$\\

\hline
Higgs\hspace{0.1cm} Doublet & $ \frac{1}{2} $&$Y=1$ &$\underline{3}$&$\phi_{i}$\\ 
\hline
Higgs\hspace{0.1cm} Doublet & $ \frac{1}{2} $&$Y=1$ &$\underline{1}$&$\phi_{0}$\\ 
\hline
Real\hspace{0.1cm} Gauge \hspace{0.1cm}Singlet  & $0$&$Y=0$ &$\underline{3}$&$F_{i}$\\ 
\hline
\end{tabular}
\end{center}
\caption{Allocations under various irreducible representations of $ SU(2)_{L} $, $ U(1)_{Y} $ and $ A_{4} $.}
\end{table}

The Yukawa Lagrangian of the leptonic fields of the model $ G_{SM} \times A_{4}\times U(1)_{X}$ {\color{blue}\cite{46}} is
\begin{equation}
\textit{L} = \textit{L}_{Charged\hspace{0.1cm}leptons\hspace{0.1cm}Dirac} + \textit{L}_{Neutrino\hspace{0.1cm}Dirac} +\textit{L}_{Neutrino\hspace{0.1cm}Majorana}
\end{equation}
where, $ G_{SM} $ is the standard model gauge symmetry,
$ G_{SM} = U(1)_{Y} \times SU(2)_{L}\times SU(3)_{C}$. Now,
\begin{equation}
\begin{split}
\textit{L}_{Charged\hspace{0.1cm}leptons\hspace{0.1cm}Dirac}  =  & - [ h_{1}
\left( \bar{Y_{1L}}\phi_{1}\right)l_{1R} +  h_{1}\left(\bar{Y_{2L}}\phi_{2}\right)l_{1R}+h_{1}\left(\bar{Y_{3L}}\phi_{3}\right)l_{1R} + h_{2}\left( \bar{Y_{1L}}\phi_{1}\right)l_{2R} \\ & +\omega^{2} \left\lbrace h_{2}\left( \bar{Y_{2L}}\phi_{2}\right)l_{2R} \right\rbrace \allowbreak + \omega \left\lbrace h_{2}\left( \bar{Y_{3L}}\phi_{3}\right)l_{2R} \right\rbrace + h_{3}\left(\bar{Y_{1L}}\phi_{1}\right)l_{3R} \\ & +\omega \left\lbrace h_{3}\left( \bar{Y_{2L}}\phi_{2}\right)l_{3R} \right\rbrace   + \omega^{2} \left\lbrace h_{3}\left( \bar{Y_{3L}}\phi_{3}\right)l_{3R} \right\rbrace ] + hc
\end{split}
\end{equation}
where,
$$ \omega = exp(\frac{2\pi i}{3}) = -\frac{1}{2} + i\frac{\sqrt{3}}{2}$$

\begin{equation}
\textit{L}_{Neutrino\hspace{0.1cm}Dirac} = - h_{0}\left(\bar{Y_{1L}}\nu_{1R}\right)\bar{\phi_{0}}- h_{0}
\left(\bar{Y_{2L}}\nu_{2R}\right)\bar{\phi_{0}} -h_{0}\left(\bar{Y_{3L}}\nu_{2R}\right)\bar{\phi_{0}} + h.c
\end{equation}
and
\begin{equation}
\begin{split}
\textit{L}_{Neutrino\hspace{0.1cm}Majorana} = & - \frac{1}{2}[\left\lbrace 
 M \nu_{1R}^{T}C^{-1}\nu_{1R} + M \nu_{2R}^{T}C^{-1}\nu_{2R} + M \nu_{3R}^{T}C^{-1}\nu_{3R}\right\rbrace  + h.c + h_{F}F_{1}\nu_{2R}^{T}C^{-1}\nu_{3R} \\ & +  h_{F}F_{1}\nu_{3R}^{T}C^{-1}\nu_{2R} +  h_{F}F_{2}\nu_{3R}^{T}C^{-1}\nu_{1R} \\ & +  h_{F}F_{2}\nu_{1R}^{T}C^{-1}\nu_{3R} +  h_{F}F_{3}\nu_{1R}^{T}C^{-1}\nu_{2R} + h_{F}F_{3}\nu_{2R}^{T}C^{-1}\nu_{1R}]
\end{split}
\end{equation}
where $ C $ is the charge conjugation matrix. $  \textit{L}_{Charged\hspace{0.1cm}leptons\hspace{0.1cm}Dirac}$ is the Dirac mass matrix in the charged leptonic fields, $ \textit{L}_{Neutrino\hspace{0.1cm}Dirac}$ is the Dirac mass matrix in the neutrino sector, $ \textit{L}_{Neutrino\hspace{0.1cm}Dirac}$ is the Dirac mass matrix in neutrino sector. 

\par 
The model here is accompanied by an additional $U(1)_{X}$ symmetry which prevents the existence of Yukawa interactions of the form $\bar{ Y_{iL}}\nu_{iR}\tilde{\phi_{i}}$ and 
$\bar{ Y_{iL}}\nu_{iR}\tilde{\phi_{0}}$ as $Y_{iL}, l_{iR}, \tilde{\phi_{0}}$ have quantum numbers $X = 1$ and all other fields have quantum numbers $X = 0$. This phenomenology disfavours Nambu Goldstone boson to arise in this case as  $U(1)_{X}$ symmetry does not break spontaneously but explicitly. Thus the Yukawa Lagrangian for the leptonic sector are of the form as described by {\color{blue} Eq. (3), (4), (5), (6)}  under the symmetry $ G_{SM} \times A_{4}\times U(1)_{X}$. 
\par
Some studies on Cosmological Domain Wall Problem are done in {\color{blue}\cite{newc}} , where it is shown that if a discrete symmetry is embedded with a continuous gauge or  global gauge group, $($in this case only $U(1)_{X}$ $)$, then on account of the phenomenon \textit{Lazarides-Shafi mechanism} the electroweak phase transition of the apparent discrete symmetry $A_{4}$  $($which is a subgroup of the centre of the continuous lie group$ ) $, results in only a network of domain walls bounded by strings to form and then quickly collapse.
\par 
Here a symmetry of the form $ U(1)_{X}$ exists. Under this symmetry $ Y_{iL} , l_{iR} $ and $ \phi_{i} $ have quantum numbers $ X=1 $ and all other fields have X = 0. This symmetry does not permit the terms like, $ \bar{Y_{L}}\nu_{R}\tilde{\phi_{i}} $ which are invariant under $ G_{SM} \times A_{4}$ and contributes to Dirac mass matrix for neutrinos. Spontaneous symmetry breaking leads to the following Vacuum expectation values for scalars, $ \upsilon_{1}, \upsilon_{2}, \upsilon_{3}$ for $ \phi_{i}'^{s} $,  $ u_{i}$ for $ F_{i}'^{s} $, $ \upsilon_{0}$ for $ \phi_{0} $. Let, $ \upsilon_{1}, \upsilon_{2}, \upsilon_{3} =\upsilon $. Here $ Y_{iL}  = (\nu_{i},l_{i}) \sim \underline{3}$, $ l_{iR}\sim \underline{1}, \underline{1^{'}}, \underline{1^{''}}$, $ \phi_{i} = (\phi_{i}^{0}, \phi_{i}^{-}) \sim \underline{3} ,(i=1,2,3)$. Along with these vacuum expectation values, the superpotential for different mass terms are:
$$ -\bar{l_{L}}M^{0}_{l}l_{R} -\bar{\nu_{L}}M_{D}\nu_{R} + \frac{1}{2}\nu_{R}^{T}C^{-1}M_{R}\nu_{R} + h.c $$ where,
\begin{equation}
M^{0}_{l} =  \begin{bmatrix}
h_{1}\upsilon_{1}& h_{2}\upsilon_{1} & h_{3}\upsilon_{1}\\
h_{1}\upsilon_{2}& h_{2}\omega^{2}\upsilon_{2} & h_{3}\omega\upsilon_{2}\\
h_{1}\upsilon_{3}& h_{2}\omega\upsilon_{3} & h_{3}\omega^{2}\upsilon_{3}\\
\end{bmatrix},M_{R} =  \begin{bmatrix}
M& h_{s}u_{3} & h_{s}u_{2}\\
h_{s}u_{3}& M & h_{s}u_{1}\\
h_{s}u_{2}& h_{s}u_{1} & M\\
\end{bmatrix}
\end{equation} 
and
$
M_{D} = h_{0}\upsilon_{0}I$. A special vacuum alignment is needed in tribimaximal 
mixing, which is given by
\begin{equation}
\upsilon_{1} = \upsilon_{2}= \upsilon_{3} = \upsilon, \hspace{0.1cm} u_{1} = u_{3}= 0, \hspace{0.1cm} \text{and} \hspace{0.1cm} h_{F}u_{2}= M^{'}.
\end{equation}
The charged lepton mass matrix $ M^{0}_{l}$ is now put in a diagonal form by the transformation,
\begin{equation}
M_{l}^{0d} = U_{\omega}M_{l}^{0}I
\end{equation}
where $ M_{l}^{0d} $ is the diagonal form of $ M_{l}^{0} $. Here
\begin{equation}
U^{C\textit{W}}_{\omega} =  \frac{1}{\sqrt{3}}\begin{bmatrix}
1& 1 & 1\\
1& \omega & \omega^{2}\\
1& \omega^{2} & \omega\\
\end{bmatrix}  \text{and}\hspace{0.1cm} \omega = exp(\frac{i2\pi}{3}) = -\frac{1}{2} + i\frac{\sqrt{3}}{2}.
\end{equation}
Now,
\begin{equation}
M_{l}^{0d}=\frac{1}{\sqrt{3}}\begin{bmatrix}
1& 1 & 1\\
1& \omega & \omega^{2}\\
1& \omega^{2} & \omega\\
\end{bmatrix}
\begin{bmatrix}
h_{1}\upsilon & h_{2}\upsilon & h_{3}\upsilon \\
h_{1}\upsilon & h_{2}\omega^{2}\upsilon & h_{3}\omega\upsilon\\
h_{1}\upsilon & h_{2}\omega\upsilon & h_{3}\omega^{2}\upsilon\\
\end{bmatrix} = \frac{1}{\sqrt{3}}\begin{bmatrix}
3h_{1}\upsilon & 0 & 0 \\
0 & 3h_{2}\upsilon & 0\\
0 & 0 & 3h_{3}\upsilon\\
\end{bmatrix}
=\begin{bmatrix}
\sqrt{3}h_{1}\upsilon & 0 & 0 \\
0 & \sqrt{3}h_{2}\upsilon & 0\\
0 & 0 & \sqrt{3}h_{3}\upsilon\\
\end{bmatrix}
\end{equation}
Or,
\begin{equation}
M_{l}^{0d}
= \begin{bmatrix}
m_{e}& 0 & 0\\
0& m_{\mu} & 0\\
0& 0& m_{\tau}\\
\end{bmatrix}.
\end{equation}
$ M_{R} $ is diagonalised by the orthogonal transformation,
\begin{equation}
U_{\nu}M_{R}U_{\nu}^{\dagger} = \begin{bmatrix} \frac{1}{\sqrt{2}} & 0 &  -\frac{1}{\sqrt{2}}\\
0 & 1 & 0\\
\frac{1}{\sqrt{2}} & 0 &  \frac{1}{\sqrt{2}}\\
\end{bmatrix}
\begin{bmatrix}
M & 0 & M^{'} \\
0 & M & 0\\
M^{'} & 0 & M\\
\end{bmatrix}
\begin{bmatrix}
\frac{1}{\sqrt{2}} & 0 &  \frac{1}{\sqrt{2}}\\
0 & 1 & 0\\
-\frac{1}{\sqrt{2}} & 0 &  \frac{1}{\sqrt{2}}\\
\end{bmatrix}
=\begin{bmatrix}
M - M^{'} & 0 & 0\\
0 & M & 0\\
0 & 0 & M + M^{'} \\
\end{bmatrix}.
\end{equation}
The choice of the vacuum alignment for scalar fields is to break $ A_{4} $ spontaneously along two incompatible directions. $ (111) $ with residual symmetry $ Z_{3} $ and $ (100) $ with residual symmetry $ Z_{2} $. The vacuum alignment breaks $ A_{4} $ in charged lepton sector coupling only with $ \phi_{i} $ to $ Z_{3} $ group. Also the vacuum alignment breaks $ A_{4} $ in neutrino sector coupling only with $ \phi_{0} $ and $F$, the residual symmetry is $ Z_{2} $ group. The PMNS matrix, tribimaximal with phases is
\begin{equation}
\frac{1}{\sqrt{3}}\begin{bmatrix} 1 & 0 &  0\\
0 & \omega & 0\\
0 & 0 &  \omega\\
\end{bmatrix}
\begin{bmatrix}
\sqrt{\frac{2}{3}} & \frac{1}{\sqrt{3}} & 0 \\
-\frac{1}{\sqrt{6}} & \frac{1}{\sqrt{3}} & -\frac{1}{\sqrt{2}}\\
-\frac{1}{\sqrt{6}} & \frac{1}{\sqrt{3}} & \frac{1}{\sqrt{2}}\\
\end{bmatrix}
\begin{bmatrix}
1 & 0& 0\\
0 & 1 & 0\\
0 & 0&- i\\
\end{bmatrix}.
\end{equation}
\section{Perturbations in Neutrino Sector}
In this section, we consider the effect of perturbations to mass matrices due to higher order corrections in the form of $ Z_{2} \times Z_{2}$ invariant perturbations. In the model discussed till now, the PMNS matrix has the tribimaximal form with zero $ \theta_{13} $ and zero $ \delta_{CP} $ phase. To generate non$-$zero values for these, small perturbation in the form of $ Z_{2} \times Z_{2}$ symmetry is added to our
model. We first instigate a symmetry breaking term in the charged lepton sector
which is invariant under the symmetry $ Z_{2} \times Z_{2}$, which is a normal subgroup of $ A_{4} $ with four elements. The three non trivial singlet representation of $ Z_{2} \times Z_{2}$ are $ \underline{\widehat{1}}^{'''}(1,1,-1,-1) $, $ \underline{\widehat{1}}^{''}(1,-1,1,-1) $, $ \underline{\widehat{1}}^{'}(1,-1,-1,1) $, whereas the one trivial singlet representation has the form $ \underline{\widehat{1}}(1,1,1,1) $. The breaking of $ A_{4} $ triplet into $ Z_{2} \times Z_{2}$ irreducible representations is given as  
\begin{equation}
(\underline{3})\hspace{0.1cm} of\hspace{0.1cm} A_{4} \longrightarrow (\underline{\widehat{1}}^{'}\oplus\underline{\widehat{1}}^{''}\oplus\underline{\widehat{1}}^{'''})
\hspace{0.1cm}of\hspace{0.1cm} Z_{2} \times Z_{2},
\end{equation}
\begin{equation}
(\underline{1}^{''},\underline{1}^{'},\underline{1})\hspace{0.1cm} of\hspace{0.1cm} A_{4} \longrightarrow (\underline{\widehat{1}})
\hspace{0.1cm}of\hspace{0.1cm} Z_{2} \times Z_{2}.
\end{equation}
If we want to break $ A_{4} $ into $ Z_{2} \times Z_{2}$ irreducible representations, it could be breaking of $ A_{4} $ triplet of right handed neutrino singlets into $ \underline{\widehat{1}}(1,1,1,1) $ trivial representations of $ Z_{2} \times Z_{2}$. 
\par 
The general $ Z_{2} \times Z_{2}$ invariant perturbations are of the form
\begin{equation}
h_{1}\bar{Y_{L}}M_{1}\phi l_{1R} + h_{2}\bar{Y_{L}}M_{2}\phi l_{2R} + h_{3}\bar{Y_{L}}M_{3}\phi l_{3R} 
\end{equation}
where, $ \bar{Y_{L}} $, $ \phi $ are the three dimensional reducible representations of $ Z_{2} \times Z_{2}$ and $ l_{R}^{'} $s are the trivial singlets. Since we have considered here $ Z_{2} \times Z_{2}$ invariant perturbations, then the matrices $ M_{1} $, $ M_{2} $, $ M_{3} $ must commute with the matrices in {\color{blue}Eq. (A11)} in Appendix A. The
$ U_{e3} $ element of the PMNS matrix in its TBM form is zero because the $11$ and $13$ elements of $ U_{\omega} $ are same. The perturbation terms in {\color{blue}Eq. (17)} can
disturb the balance between the $11$ and $13$ elements of $ U_{\omega} $ and this phenomenology leads to non$-$zero $ \theta_{13} $. The value of $ \theta_{13} $ relates to the elements of the mass matrices, $ M_{1} $, $ M_{2} $, $ M_{3} $. We prefer the form of $ M_{i}$'s as $M_{i} = diag(\bar{z},0,\omega^{i-1}z)$ to generate simple form of perturbed charged lepton mass matrices $ M_{l} $. $ z $ is a complex number and $ |z|<1 $. After spontaneous symmetry breaking the resulting $ M_{l} = M_{l}^{0} + \delta M_{l}$ where, $ M_{l}^{0}$ has the form in  {\color{blue}Eq. (7)}, and $\delta M_{l}$ has the form 
\begin{equation}
\delta M_{l} = \begin{bmatrix}
h_{1}\upsilon\bar{z} & h_{2}\upsilon\bar{z} & h_{3}\upsilon\bar{z} \\
0 & 0 & 0\\
h_{1}\upsilon z & h_{2}\upsilon z \omega & h_{3}\upsilon z \omega^{2}\\
\end{bmatrix}.
\end{equation}
$ \delta M_{l} $ arises from the higher order effects of the theory. We parameterize all the higher order perturbations in the terms of the complex number $ z $ {\color{blue}\cite{39c}}. There is no residual symmetry remaining in the charged lepton sector after the spontaneous symmetry breaking. Constraining $ U_{\omega} $ to be unitary, we limit $ z $ as 
\begin{equation}
z = -1 \pm \sqrt{1-S^{2}} + iS.
\end{equation}
For  $ z<1 $ one gets perturbation to be of the order $ S $. Using the parametrization
 $ S= Sin\hspace{0.1cm}\alpha $, $ U_{\omega} $ to 
\begin{equation}
U_{\omega} = \frac{1}{\sqrt{3}} \begin{bmatrix}
e^{i \alpha} & 1& e^{-i \alpha}\\
e^{i \alpha} & \omega& \omega^{2}e^{-i \alpha}\\
e^{i \alpha} & \omega^{2} &\omega e^{-i \alpha}\\
\end{bmatrix},
\end{equation}
we introduce a $ Z_{2} \times Z_{2}$ invariant perturbations in the neutrino sector, and study its influence on $ \theta_{13} $ and $ \delta_{CP} $. The perturbing matrix is diagonal since it should satisfy $ Z_{2} \times Z_{2}$ symmetry. We chose the perturbation {\color{blue}\cite{50}} to be as follows:
$$M\nu_{R}^{T}C^{-1}\begin{bmatrix}
\frac{1}{\rho} e^-{i\varphi} & 0& 0\\
0 & 0 & 0\\
0 & 0&\frac{1}{\rho} e^-{i\varphi}\\
\end{bmatrix}\nu_{R}$$
where, $\frac{1}{\rho} e^-{i\varphi}$ characterises the soft brreaking of $ A_{4} $. M is $ A_{4} $ invariant soft term in the Lagrangian. The perturbing term is $ A_{4} $ breaking but $ Z_{2} \times Z_{2}$ invariant soft term in the Lagrangian. The perturbed matrix is now
\begin{equation}
\begin{bmatrix}
M+\frac{1}{\rho} e^-{i\varphi}M & 0& M^{'}\\
0 & M & 0\\
M^{'} & 0 & M-\frac{1}{\rho} e^-{i\varphi}M\\
\end{bmatrix}.
\end{equation} 
We can diagonalise it by rotation angle $ x $, where, 
\begin{equation}
Tan \hspace{0.1cm}2x = \frac{M^{'}}{\frac{1}{\rho} e^-{i\varphi}M}.
\end{equation} 
Thus, we see that the ratio, $ \frac{M^{'}}{M} $ is a physical observable in the rotation angle, $Tan \hspace{0.1cm}2x$ which helps us to diagonalise the perturbing matrix {\color{blue}Eq. (21)}. The introduction of the perturbing terms like $\frac{1}{\rho} e^-{i\varphi}$ and the ratio, $ \frac{M^{'}}{M} $ has helped us to derive non zero $ \theta_{13} $ and other neutrino oscillation parameters in terms of $ x, \alpha, \rho  $, {\color{blue}Eq. (24)-(37)}. The input range of Majorana phases used in our calculation is from 0 to 2$ \pi$. The heavy right handed Majorana neutrino used here for the computation of various LFV decay rates like $ \mu\rightarrow e\gamma $, $ \tau\rightarrow \mu \gamma $, $ \tau\rightarrow e\gamma $ is $ 10^{15} $ GeV.
\par
The most interesting feature of our work is that, from {\color{blue}Eq. (24)-(37)} we can extract meaningful extract of current pattern of neutrino flavour mixing in the sense that the favoured value of $ \delta_{CP} $ phase from our results attached with 
non$-$zero $\theta_{13}$ can induce signatures of various decay rates of charged lepton flavour violation processes like  $ \mu\rightarrow e\gamma $, $ \tau\rightarrow \mu \gamma $, $ \tau\rightarrow e\gamma $ after spontaneous symmetry breaking of our model  $ G_{SM} \times A_{4}\times U(1)_{X}$ incorporating $ Z_{2} \times Z_{2}$ invariant perturbations into account. The sleptons and gauginos so constrained are shown in 
{\color{blue}Figs. 11-14}. The prospect to test these sparticles at future run of LHC will favour or rule out our model.
\par
After introducing $ Z_{2} \times Z_{2}$ perturbations in both charged leptonic sector and neutrino sector {\color{blue}\cite{50,ra,pat}}, and setting $U_{\omega}$ as defined by {\color{blue}Eq. (20)}, PMNS matrix after perturbation becomes 
\begin{equation}
U_{PMNS}  = \frac{1}{\sqrt{3}}\begin{bmatrix}
e^{i \alpha} & 1& e^{-i \alpha}\\
e^{i \alpha} & \omega& \omega^{2}e^{-i \alpha}\\
e^{i \alpha} & \omega^{2} &\omega e^{-i \alpha}\\
\end{bmatrix}\begin{bmatrix}
Cos x & 0 & -Sin x\\
0 & 1 & 0\\
Sin x & 0 & Cos x\\
\end{bmatrix}.
\end{equation}
From above it is seen that, after computation matching with the actual PMNS matix one gets,
\begin{equation}
\text{Sin}\theta_{13}e^{-i\delta_{CP}} = \frac{1}{\sqrt{3}}\left( e^{-i\alpha}\text{Cosx}-e^{i\alpha}\text{Sinx}\right)
=  \frac{1}{\sqrt{3}}\left( \text{Cos}\alpha \text{Cosx}-\text{Cos}\alpha \text{Sinx} - i\text{Sin}\alpha \text{Cosx}- i\text{Sin}\alpha \text{Sinx}\right). 
\end{equation}
Therefore,
\begin{equation}
\text{Sin}\theta_{13}Cos\delta_{CP} = \frac{1}{\sqrt{3}}\left( \text{Cos}\alpha \text{Cosx}-\text{Cos}\alpha \text{Sinx}\right),
 \end{equation}
 and
 \begin{equation}
\text{Sin}\theta_{13}Sin\delta_{CP} = \frac{1}{\sqrt{3}}\left( \text{Sin}\alpha \text{Cosx}+\text{Sin}\alpha \text{Sinx}\right).
 \end{equation}
Squaring and adding {\color{blue}Eq. (25)} and {\color{blue}Eq. (26)} we get
\begin{equation}
\begin{split}
Sin^{2}\theta_{13}(Cos^{2}\delta_{CP} + Sin^{2}\delta_{CP})  = \frac{1}{( \sqrt{3}) ^{2}}\left\lbrace \left(Cos\hspace{0.1cm} \alpha\hspace{0.1cm}Cos\hspace{0.1cm}x-Cos\hspace{0.1cm}  \alpha\hspace{0.1cm}Sin\hspace{0.1cm}x\right)^{2}+\left(Sin\hspace{0.1cm} \alpha\hspace{0.1cm}Cos\hspace{0.1cm}x+Sin\hspace{0.1cm}  \alpha\hspace{0.1cm}Cos\hspace{0.1cm}x\right)^{2}\right\rbrace \\
=  \frac{1}{3}\left\lbrace Cos\hspace{0.1cm}^{2}\alpha\hspace{0.1cm}Cos\hspace{0.1cm}^{2}x +Cos\hspace{0.1cm}^{2}\alpha\hspace{0.1cm}Sin\hspace{0.1cm}^{2}x - 2Cos\hspace{0.1cm}^{2}\alpha Cos\hspace{0.1cm}x\hspace{0.1cm}Sin\hspace{0.1cm}x + Sin\hspace{0.1cm}^{2}\alpha\hspace{0.1cm}Cos\hspace{0.1cm}^{2}x +Sin\hspace{0.1cm}^{2}\alpha\hspace{0.1cm}Sin\hspace{0.1cm}^{2}x + 2Sin^{2}\hspace{0.1cm}\alpha Cos\hspace{0.1cm}x\hspace{0.1cm}Sin\hspace{0.1cm}x\right\rbrace  \\
=\frac{1}{3}\left\lbrace 1-Cos\hspace{0.1cm}2\alpha\hspace{0.1cm}Sin\hspace{0.1cm}2x \right\rbrace  
=\frac{1}{3}\left\lbrace 1-Sin\hspace{0.1cm}2x(1-2Sin\hspace{0.1cm}^{2}\alpha) \right\rbrace 
=\frac{1}{3}\left\lbrace 1-Sin\hspace{0.1cm}2x(1-2S^{2}) \right\rbrace \\
=\frac{1}{3}\left\lbrace 1-\frac{1}{\sqrt{1+Cot\hspace{0.1cm}^{2}2x}}(1-2S^{2}) \right\rbrace
= \frac{1}{3}\left\lbrace 1-\frac{1}{\sqrt{1+\kappa^{2}}}(1-2S^{2}) \right\rbrace
=\frac{1}{3}\left\lbrace 1-(1+\frac{-1}{2}\kappa^{2})(1-2S^{2}) \right\rbrace 
= \frac{\kappa^{2}}{6} + \frac{2}{3}S^{2} - \frac{\kappa^{2}S^{2}}{3}\\
\end{split},
\end{equation}
or
\begin{equation}
Sin^{2}\theta_{13} = \frac{1}{3}\left( 1-Cos\hspace{0.1cm} 2 \alpha\hspace{0.1cm} Sin\hspace{0.1cm} 2x\right) 
=  \frac{\kappa^{2}}{6} + \frac{2}{3}S^{2} - \frac{\kappa^{2}S^{2}}{3},
\end{equation} 
where perturbations in neutrino sector is defined by 
\begin{equation}
\kappa =  \frac{1}{\rho} e^-{i\varphi}\frac{M}{M^{'}} = Cot 2x.
\end{equation} 
and $ S = Sin \hspace{0.1cm} \alpha$.
Therefore, in terms of soft breaking parameters, one gets 
\begin{equation}
Sin^{2}\theta_{13} 
=  \frac{1}{6\rho^{2}} e^-{2i\varphi}\frac{M^{2}}{M^{'2}} + \frac{2}{3}Sin^{2} \hspace{0.1cm} \alpha - \frac{1}{3\rho^{2}} e^-{2i\varphi}{\frac{M^{2}}{M^{'2}}}^{2}Sin^{2} \hspace{0.1cm} \alpha.
\end{equation} 
Similarly one has,
\begin{equation}
Sin^{2}\hspace{0.1cm}\theta_{12}Cos^{2}\hspace{0.1cm}\theta_{13} = \frac{1}{3},
\end{equation}
and also
\begin{equation}
Cos^{2}\hspace{0.1cm}\theta_{13} = \frac{1}{3}(2 + Sin\hspace{0.1cm}2xCos\hspace{0.1cm}2\alpha).
\end{equation}
Thus, 
\begin{equation}
Sin^{2}\hspace{0.1cm}\theta_{12} = \frac{1}{2 + Sin\hspace{0.1cm}2xCos\hspace{0.1cm}2\alpha}.
\end{equation}
Inflating the above expression for $Sin^{2}\hspace{0.1cm}\theta_{12} $ upto the order $ \kappa^{2} $ and $S^{2}$, one gets
\begin{equation}
Sin^{2}\hspace{0.1cm}\theta_{12} = \frac{1}{3} + \frac{2}{9}S^{2} + 
\frac{\kappa^{2}}{18}-\frac{\kappa^{2}S^{2}}{27} 
=  \frac{1}{3} + \frac{2}{9}Sin^{2}\alpha + 
\frac{1}{18\rho^{2}}e^-{2i\varphi\frac{M^{2}}{M^{'2}}}-\frac{1}{27\rho^{2}} e^-{2i\varphi\frac{M^{2}}{M^{'2}}Sin^{2}\alpha}.
\end{equation}
Similarly we get in terms of soft breaking parameters, after computation matching with the actual PMNS matix,
\begin{equation}
Sin^{2}\hspace{0.1cm}\theta_{23} = \frac{\sqrt{3}Sin2xSin2\alpha + 2 + Sin2xCos2\alpha}{4 + Sin2xCos2\alpha}
=  0.5 + \frac{Sin\hspace{0.1cm}\alpha}{\sqrt{3}} - 
\frac{1}{3\sqrt{3}\rho^{2}}e^-{2i\varphi\frac{M^{2}}{M^{'2}}}Sin\hspace{0.1cm}\alpha.
\end{equation}
Finally, we have
\begin{equation}
Cos\hspace{0.1cm}\delta_{CP} = \sqrt{1-\frac{Cos^{2}\hspace{0.1cm}2x\left( 2+Cos\hspace{0.1cm} 2 \alpha\hspace{0.1cm} Sin\hspace{0.1cm} 2x\right)^{2}}{\left(  1-Cos^{2}\hspace{0.1cm} 2 \alpha\hspace{0.1cm} Sin^{2}\hspace{0.1cm} 2x\right) \left[ 4 + 4 Cos\hspace{0.1cm} 2 \alpha\hspace{0.1cm} Sin\hspace{0.1cm} 2x +(-1+2 Cos 4\alpha)Sin^{2}2x\right]}}
\end{equation} 
or
\begin{equation}
Cos\hspace{0.1cm}\delta_{CP} = \sqrt{1-\frac{(-\kappa)^{2}}{4Sin^{2}\alpha + \kappa^{2}-16\frac{\kappa^{2}Sin^{2}\alpha}{3}}}
=  \sqrt{1-\frac{\frac{1}{\rho^{2}} e^-{2i\varphi}\frac{M^{2}}{M^{'2}}^{2}}{4Sin^{2}\alpha + \frac{1}{\rho^{2}} e^-{2i\varphi}\frac{M^{2}}{M^{'2}}^{2}-16\frac{\frac{1}{\rho^{2}} e^-{2i\varphi}\frac{M^{2}}{M^{'2}}^{2}Sin^{2}\alpha}{3}}},
\end{equation}
keeping the leading powers in numerators and denominator. For no perturbation in neutrino sector, the value of $ \delta_{CP} $ becomes zero as $ \kappa $ tends to zero. Owing to perturbation inhibition only in neutrino sector, one sets $ S=0 $ or $ \delta_{CP}= \pm \frac{\pi}{2}$. In conformity with T2k data from the $ \nu_{e} $ appearance data, the value of $ \delta_{CP} $ is favoured to be in the lower half plane.
In the wake of perturbations only in the neutrino sector, one gets
\begin{equation}
Sin\hspace{0.1cm}\delta_{CP} = -Sin\frac{\pi}{2}.
\end{equation}
This implies $S\longrightarrow 0 $ or $ \kappa $ is positive and perturbations in charged leptonic sector is imperceptible. 
\par
As we work in the type I seesaw framework, the heavy Majorana neutrino mass scale M is  much higher than the electroweak scale and the light neutrinos are simply given by the well-known effective mass matrix
\begin{equation}
m_{\nu} = -M_{D}M_{R}^{-1}m_{D}^{T},
\end{equation}
where $ M_{D} $ is the Dirac $-$ type neutrino mass matrix in the weak basis. The charged lepton masses are real and diagonal.
\begin{equation}
M_{D} = \upsilon_{0}U_{\omega}^{\dagger}Y_{\nu},
\end{equation}
where $\upsilon_{0}$ denotes the vacuum expectation value of the usual SM Higgs doublet, $ <\phi_{0}> =\upsilon_{0}$. We thus compute the form of $ Y_{\nu} $, the Dirac neutrino yukawa couplings (DNY) from $
M_{D} = h_{0}\upsilon_{0}I$ and {\color{blue}Eq. (40)}, after spontaneous breaking of $ A_{4} $ symmetry incorporating $ Z_{2} \times Z_{2}$ invariant peturbations in the neutrino sector and thus generating non zero $ \theta_{13} $ and $ \delta_{CP} $.
\begin{equation}
Y_{\nu} = h_{0}|U_{\omega}^{\dagger}|^{-1}.
\end{equation}
Here $h_{0}$ is the arbitrary coupling constant giving the lepton masses. Taking $h_{0}$ of the order of 0.08 we can construct a user defined neutrino Yukawa coupling $Y_{\nu}$ corresponding to favoured value of $ \alpha $ from our results.
\par 
Charged lepton flavour violation (CLFV) and hence neutrino oscillations and mixings are real phenomenon. In terms of low energy observables, the lepton flavour violating entries in the SO(10) SUSY GUT framework can be expressed as
\begin{equation}
 \left( m^{2}_{\tilde{L}} \right)_{i\neq j} = \frac{-3m_{o}^{2}+A_{o}^{2}}{ 8\pi^{2}} \sum_{k}
      \left(Y_{\nu}^{\star}\right)_{ik}\left(Y_{\nu}\right)_{jk} \\ log\left(\frac{M_{X}}{M_{R_{k}}}\right), 
\end{equation}
 here $M_{X}$ is the GUT scale, $M_{R_{k}}$ is the  scale of the $k^{th}$ heavy RH majorana neutrino,  $m_{0}$ and $A_{0}$ are universal soft mass and trilinear terms at the high scale. $ Y_{\nu} $ are the Dirac neutrino Yukawa couplings. The flavor violating off-diagonal entries at the weak scale are found by using $Y_{\nu}$. The branching ratio of a charged lepton flavour violating decay {\color{blue}\cite{GG}} $ l_{i}$ $\rightarrow $ $l_{j}$ is 
\begin{equation}
\text{BR} \left( l_{i} \rightarrow l_{j}+\gamma \right)\approx \alpha^{3}\frac{\vert\delta^{LL}_{ij}\vert^{2}}{G_{F}^{2}M^{4}_{SUSY}}\hspace{0.1cm} Tan^{2}\hspace{.01cm}\beta \linebreak\hspace{0.1cm} \text{BR} \left( l_{i}  \rightarrow  l_{j}\nu_{i}\tilde{\nu_{j}} \right).
\end{equation}
The most interesting feature of this work is that we predicted form of Dirac neutrino Yukawa coupling $Y_{\nu}$ corresponding to favoured value of 
$ \alpha \sim 60^{0} $ which could realise signatures of rare CLFV decays like  $ \mu\rightarrow e\gamma $, $ \tau\rightarrow \mu \gamma $, $ \tau\rightarrow e\gamma $  after spontaneous symmetry breaking of our model $ G_{SM} \times A_{4}\times U(1)_{X}$ incorporating $ Z_{2} \times Z_{2}$ invariant perturbations into account. In this context we used the value of Higgs mass as measured at LHC, latest global data on the reactor mixing angle $ \theta_{13} $ for neutrinos, and latest constraints on BR($ \mu  \rightarrow e  \gamma $) as projected by MEG at PSI and MEG II PSI {\color{blue}\cite{14nov}} planning to achieve sensitivity to BR$ (\mu\rightarrow e\gamma)\sim 10 ^{-14}$. Beyond Standard Model Physics that could yield CLFV embraces SUSY sparticles, $ Z^{'} $ vector bosons with flavour non diagonal couplings which is an indication of lepton flavour violation.
\par 
Signatures of CLFV could be tested by the late 2020s at next run of High $-$ Luminosity LHC, if SUSY sparticles are observed within few TeV range, as discussed in detail below. It is worth mentioning here that, during last run of LHC, no SUSY partner of SM has been observed, and this could point to a high scale SUSY theory. Split SUSY offers a dark matter candidate and unifies the fundamental forces at high energies, it doesn't address the stability of Higgs boson . 
\begin{center}
\begin{figure*}[htbp]
\centering{
\begin{subfigure}[]{\includegraphics[height=8cm,width=8.99cm]{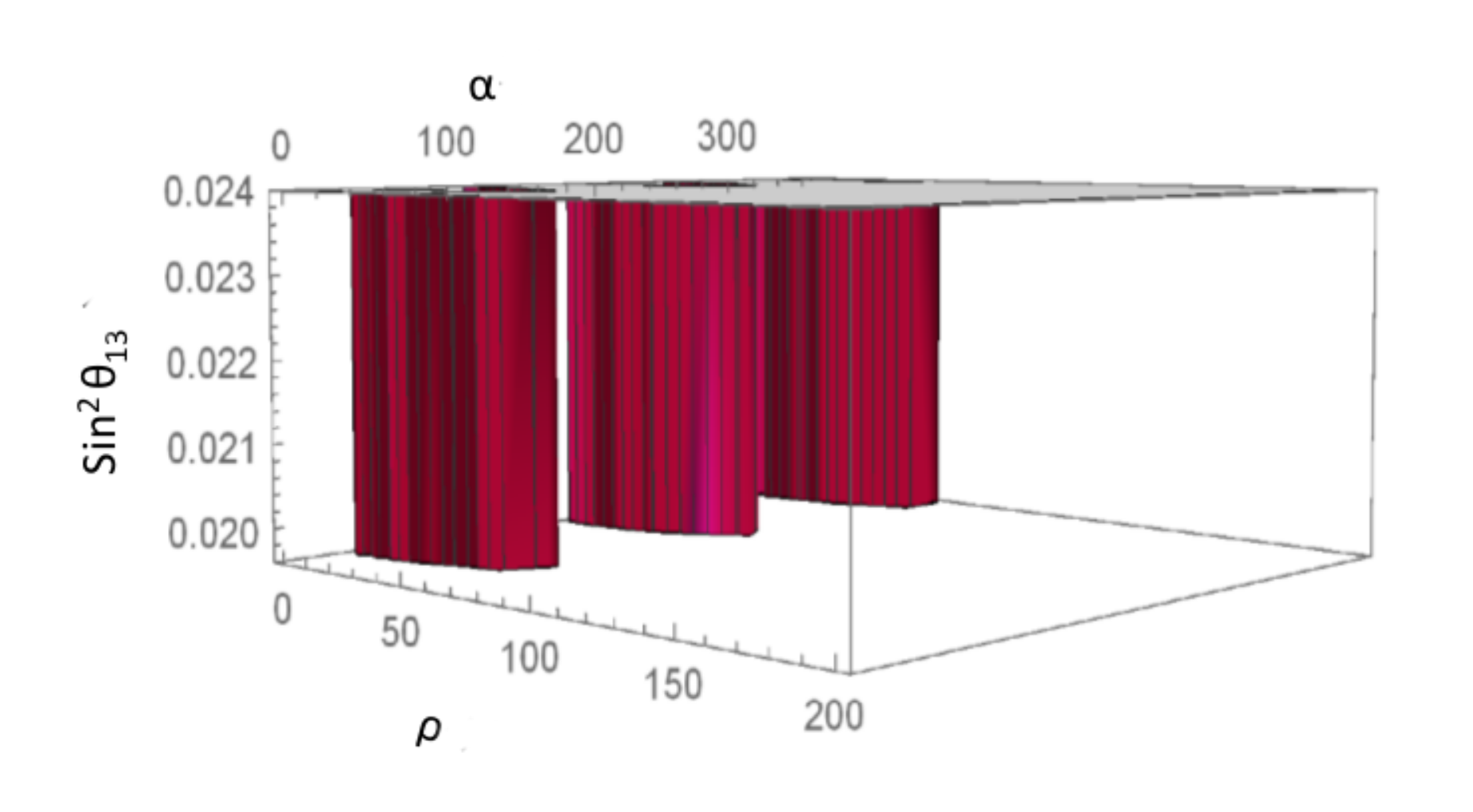}}\end{subfigure}
\begin{subfigure}[]{\includegraphics[height=	8cm,width=8.8cm]{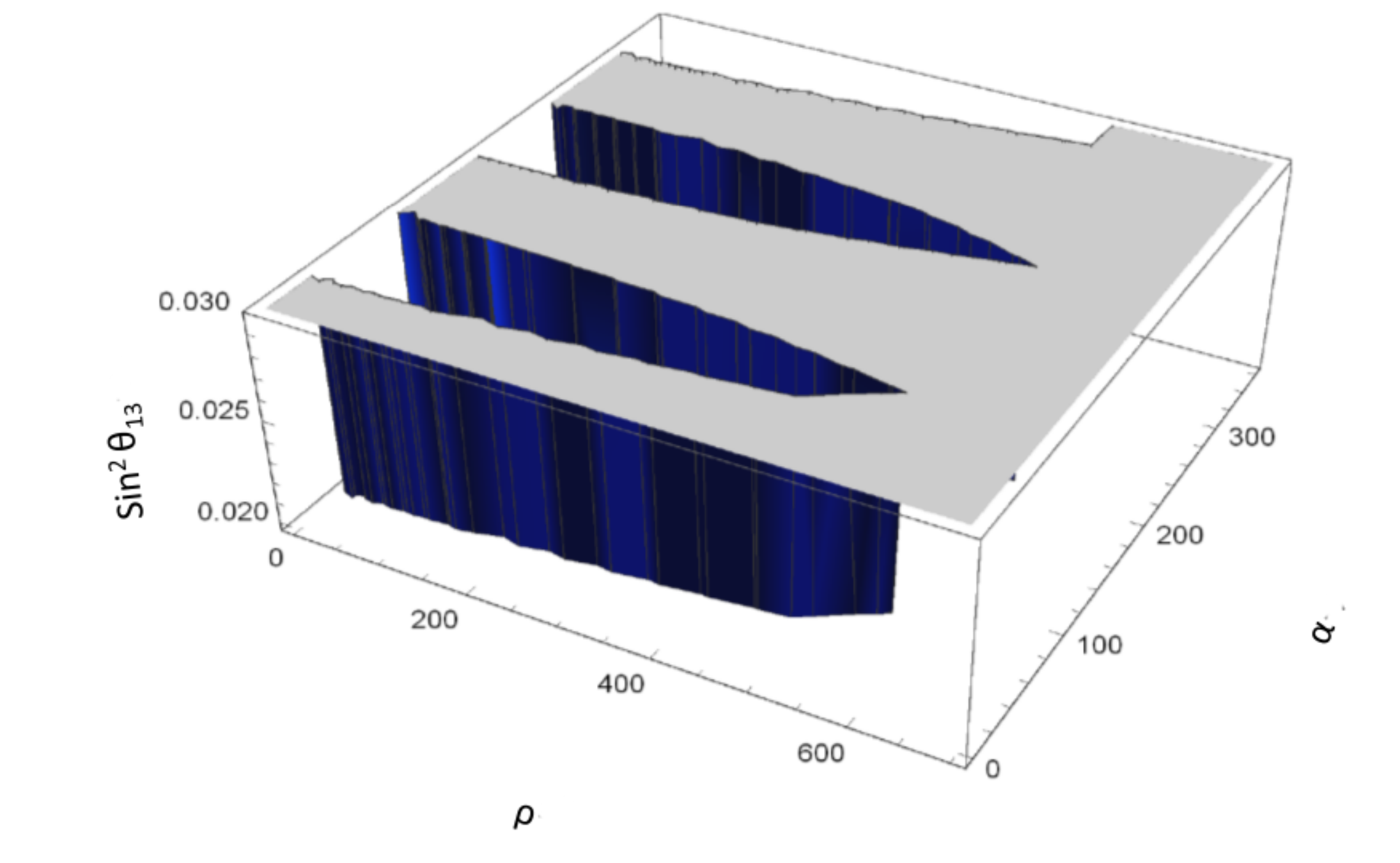}}\end{subfigure}\\
\caption{In {\color{blue}Fig. 1(a)} the points in $\rho-\alpha $ space which satisfy the 3$ \sigma $ constraints on $Sin^{2}\theta_{13}$ for $ \frac{M^{'}}{M} =10^{-2}$ are presented. {\color{blue}Fig. 1(b)} shows the points in $Sin^{2}\theta_{13}-\rho-\alpha $ space corresponding to  the 3$ \sigma $ bounds on $Sin^{2}\theta_{13}$ for $ \frac{M^{'}}{M}=10^{-3}$.}} 
\end{figure*}
\end{center}
To show that the model predicts the neutrino mixing angles compatible with the observed data, we obtain the allowed parameter space for the correction terms in the perturbation matrix in the form of $ Z_{2} \times Z_{2}$ invariant symmetry compatible with the 3$\sigma$ range of the observed $ \nu $ oscillation data by varying the parameters. Considering, definite values for $ \frac{M^{'}}{M} =10^{-2},10^{-3}$ and setting forth the values for soft breaking phase as $ \varphi = 0,\pm\frac{\pi}{2},\pi$, we generate the soft breaking parameter space for $ \rho $ and  $ \alpha $ from the $3\sigma$ constraints on mixing angles, $ Sin^{2}\theta_{12}$, $ Sin^{2}\theta_{23}$, $ Sin^{2}\theta_{13}$, and  CP violating Phase, $ \delta_{CP} $. As can be seen from {\color{blue}Eq. (30)} that to generate non-zero values of $ \theta_{13}$, soft breaking phase $\varphi$ should take the values as $0,\pm\frac{\pi}{2},\pi$ as $\varphi$ is present in the form of $\frac{1}{6\rho^{2}} e^-{2i\varphi}$ in the expression of $\theta_{13}$. $e^-{2i\varphi}$ will take real values only for $\varphi = 0,\pm\frac{\pi}{2},\pi$, since, other values of $\varphi$, such as $\pm\frac{\pi}{4}$ will generate imaginary values of  $\theta_{13}$ in the form of $ iSin 2 \varphi $ which is absurd. Similarly {\color{blue}Eq. (34)}, {\color{blue}Eq. (35)} and {\color{blue}Eq. (37)} will take real favoured values of mixing angles $\theta_{12}$, $\theta_{23}$ and CP violating phase $\delta_{CP}$ respectively for $\varphi = 0,\pm\frac{\pi}{2},\pi$. 
\par 
Owing to the {\color{blue}Eq. (37)} one finds that the value of $ \delta_{CP} $ depends on the comparative supremacy between the parameters $ Sin\hspace{0.1cm}\alpha $ and $ \kappa $ or $ \rho$. This obsession between the parameters $ Sin\hspace{0.1cm}\alpha $ and $ \kappa $ or $ \rho$ in procuring the mixing angles and CP violation phase, $ \delta_{CP} $ in terms of {\color{blue}Eqs. (28), (29), (30), (34), (35), (37)} by virtue of perturbations in neutrino sector and charged leptonic sector has been plotted in {\color{blue}Figs. 1-10}.

\begin{center}
\begin{figure*}[htbp]
\centering{
\begin{subfigure}[]{\includegraphics[height=8.8cm,width=8.8cm]{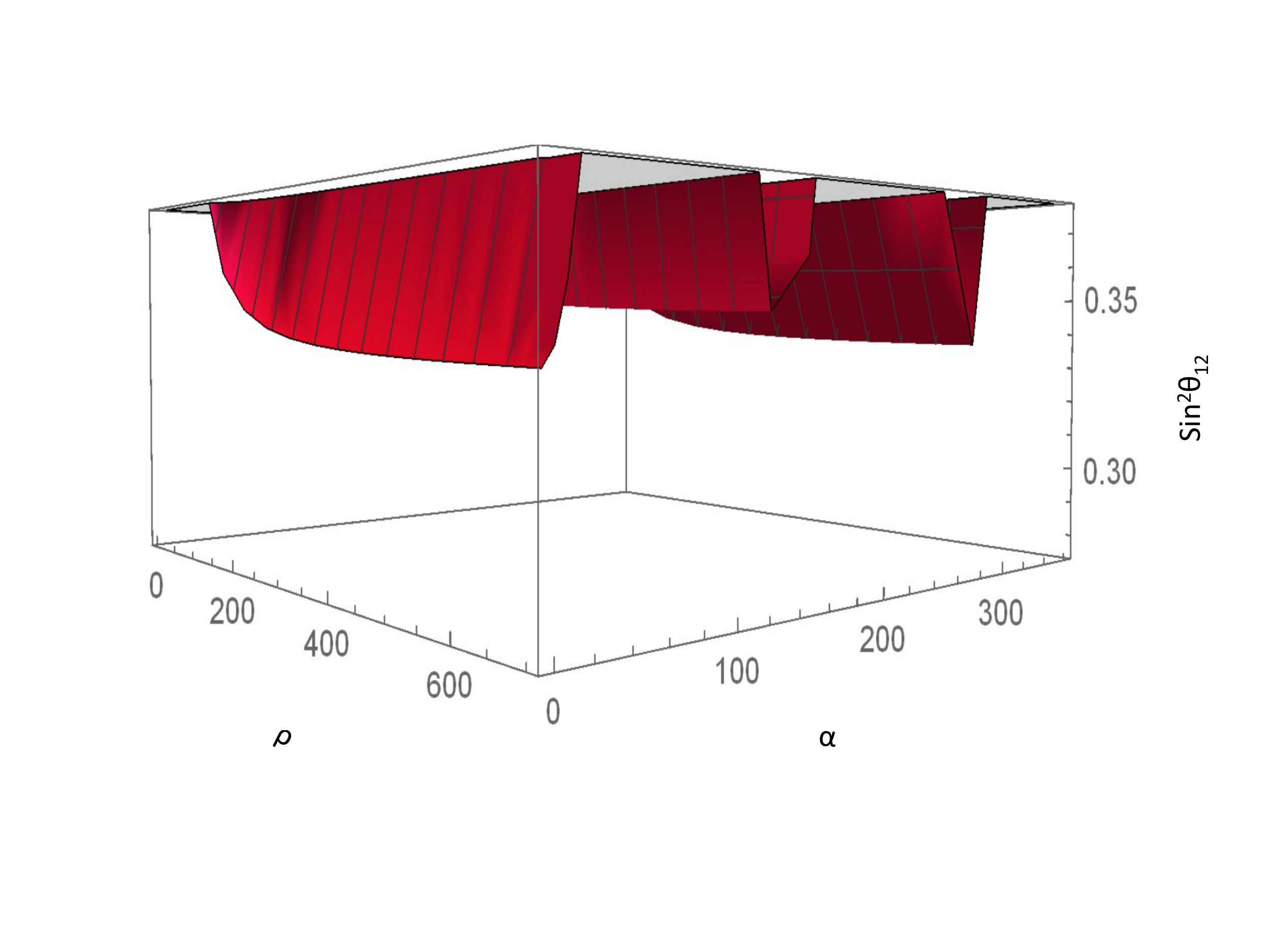}}\end{subfigure}
\begin{subfigure}[]{\includegraphics[height=	8.4cm,width=8.8cm]{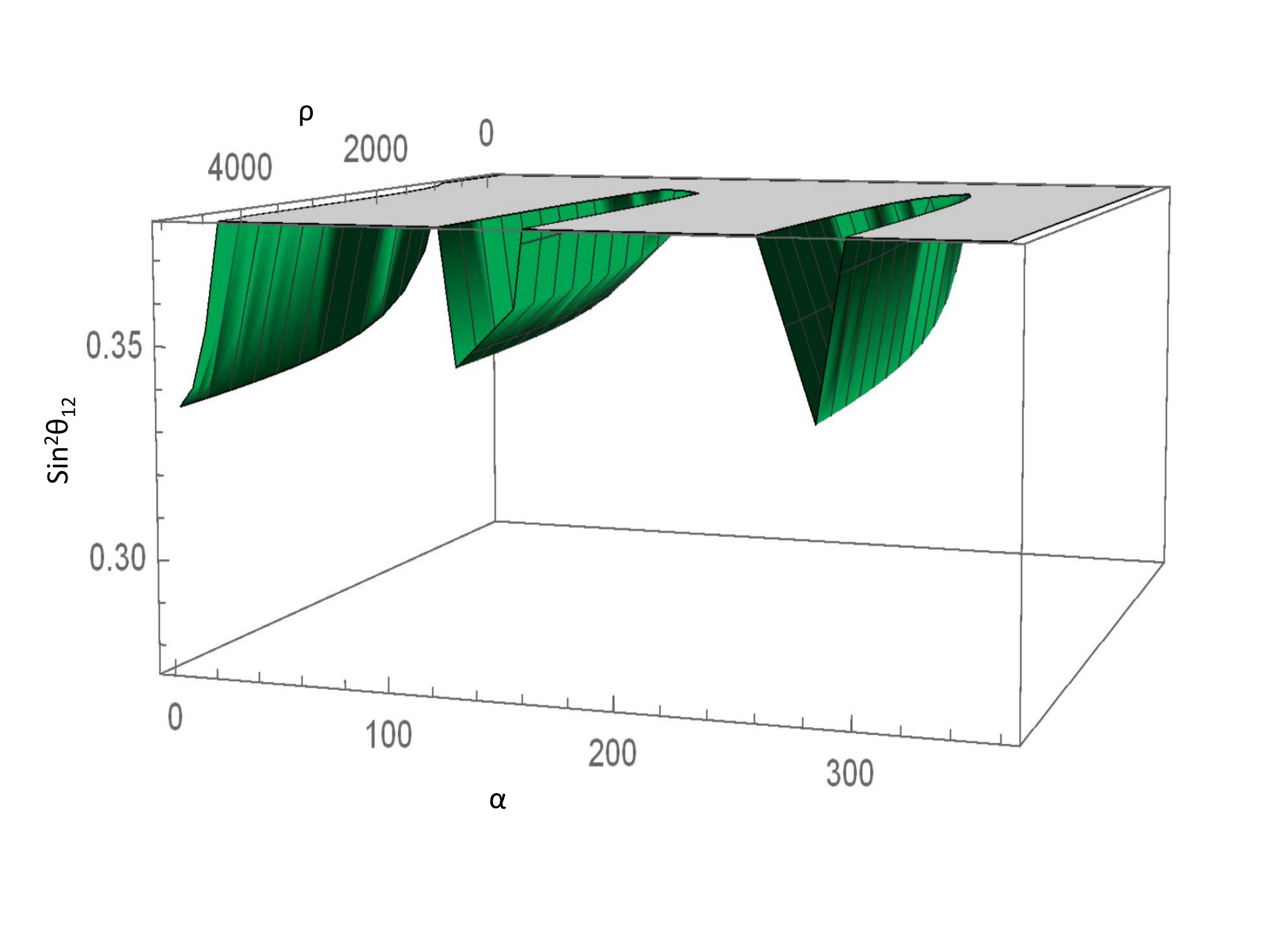}}\end{subfigure}\\
\caption{In {\color{blue}Fig. 2(a)} the allowed range of $\rho-\alpha $ space which satisfy the 3$ \sigma $ constraints on $Sin^{2}\theta_{12}$ for $ \frac{M^{'}}{M} =10^{-2}$ are shown. {\color{blue}Fig. 2(a)} shows the correlation between $\rho$ and $\alpha$ space corresponding to the 3$ \sigma $ bounds on $Sin^{2}\theta_{12}$ for $ \frac{M^{'}}{M}=10^{-3}$.}}. 
\label{fig:1}
\end{figure*}
\end{center}
In {\color{blue}Fig. 1} the predicted dependence of $Sin^{2}\theta_{13}$ on the soft breaking parameter, $ \rho $ and  $ \alpha $ is shown. Owing to the constrained nature of  the mixing angle, $Sin^{2}\theta_{13}$ varying within its 3$ \sigma $ range indicated by current neutrino oscillation global fit {\color{blue}\cite{pc}}, one finds the correlation between mixing angle, $Sin^{2}\theta_{13}$, and $ \rho $, $ \alpha $ breaking parameter space corresponding to $\frac{M^{'}}{M}=10^{-2}$ and $ \frac{M^{'}}{M}=10^{-3}$ in the left and right panel respectively. Owing to $ Sin\hspace{0.1cm}\alpha $ and $ Z_{2} \times Z_{2}$ perturbations in the neutrino sector the allowed range of $ \rho $  space lies in the range $[0,50] $ and $[0,560]$ for $ \frac{M^{'}}{M}=10^{-2}$ and $ \frac{M^{'}}{M}=10^{-3}$ respectively. {\color{blue}Fig. 2}, shows the variation of $Sin^{2}\theta_{12}$ within its 3 $ \sigma $ range with respect to the soft breaking parameter, $ \rho $ and $ \alpha $. The curtailment of the mixing angle, $Sin^{2}\theta_{12}$ differing within its 3$ \sigma $ range allowed by the current neutrino oscillation global fit, one finds the whole experimentally allowed range of $ \rho $, $ \alpha $ breaking parameter space corresponding to $\frac{M^{'}}{M}=10^{-2}$ and $ \frac{M^{'}}{M}=10^{-3}$ in the left and right panel respectively. On account of $ Sin\hspace{0.1cm}\alpha $ and $ Z_{2} \times Z_{2}$ perturbations in the neutrino sector the predicted range of $ \rho $  space lies in the range $[0,600] $ and $[1500,4500]$ for $ \frac{M^{'}}{M}=10^{-2}$ and $ \frac{M^{'}}{M}=10^{-3}$ respectively. From {\color{blue}Fig. 1}, one finds that the value of $ Sin^{2}\theta_{13} = 0.0216$ is near $ \rho = 560 $. For this value of $ \rho $, the change in $ Sin^{2}\theta_{12} $ is comparatively small $ (\sim 6) \%$. Thus, one finds that the value of $ \rho $ is comparatively large but the parameter $\rho e^-{i\varphi} = \kappa\frac{M^{'}}{M}$ appraising the perturbation in the neutrino sector is utterly small because $ \frac{M^{'}}{M} <<1$ and also owing to the contributions from $ \frac{1}{\rho} $ factor. In the limiting case, $ Sin\hspace{0.1cm}\alpha\rightarrow 0 $, the value of $ Sin^{2}\theta_{23} $ becomes $ \frac{1}{2} $. 
\begin{center}
\begin{figure*}[htbp]
\centering{
\begin{subfigure}[]{\includegraphics[height=8cm,width=8.7cm]{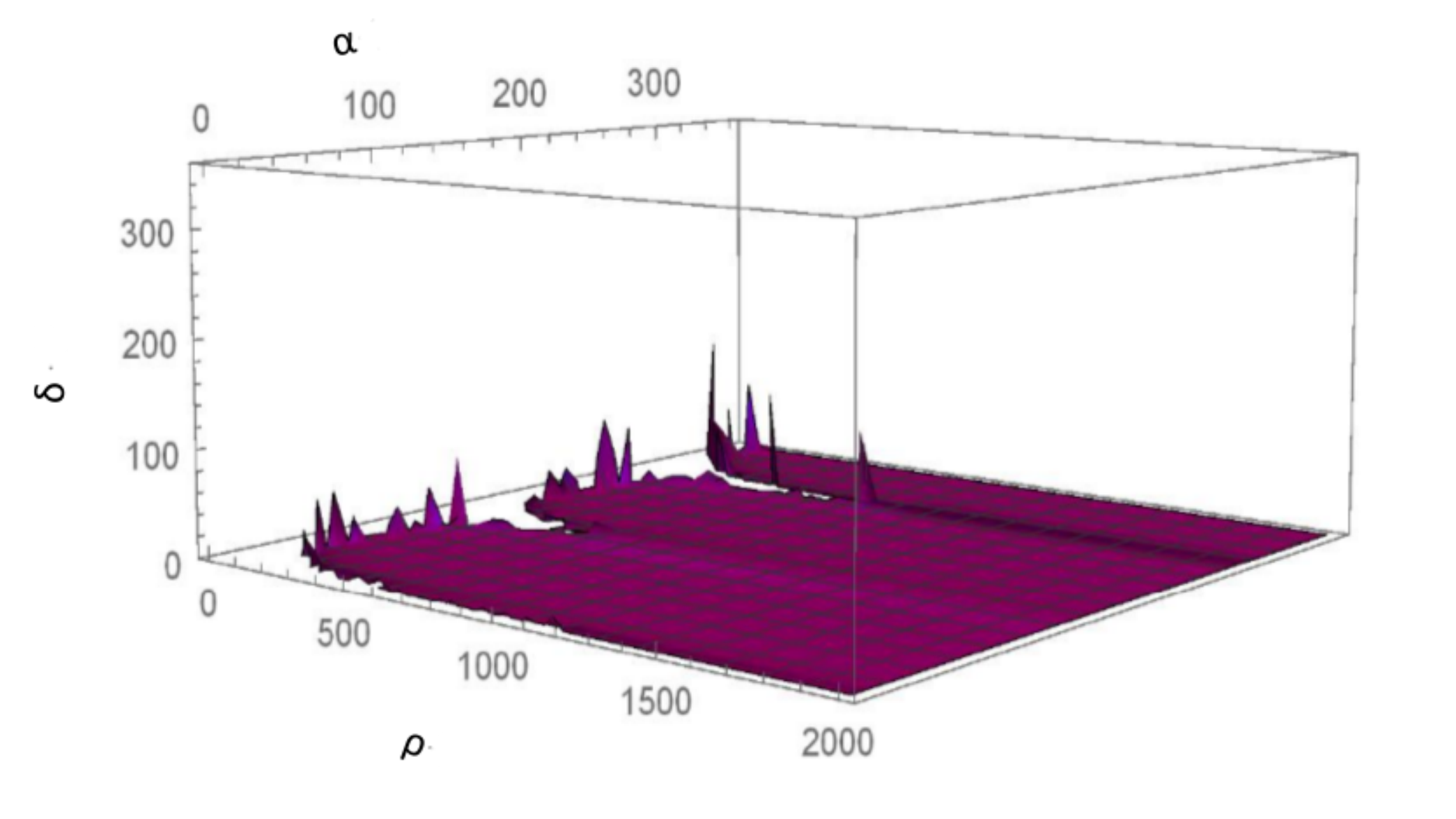}}\end{subfigure}
\begin{subfigure}[]{\includegraphics[height=	8.2cm,width=8.7cm]{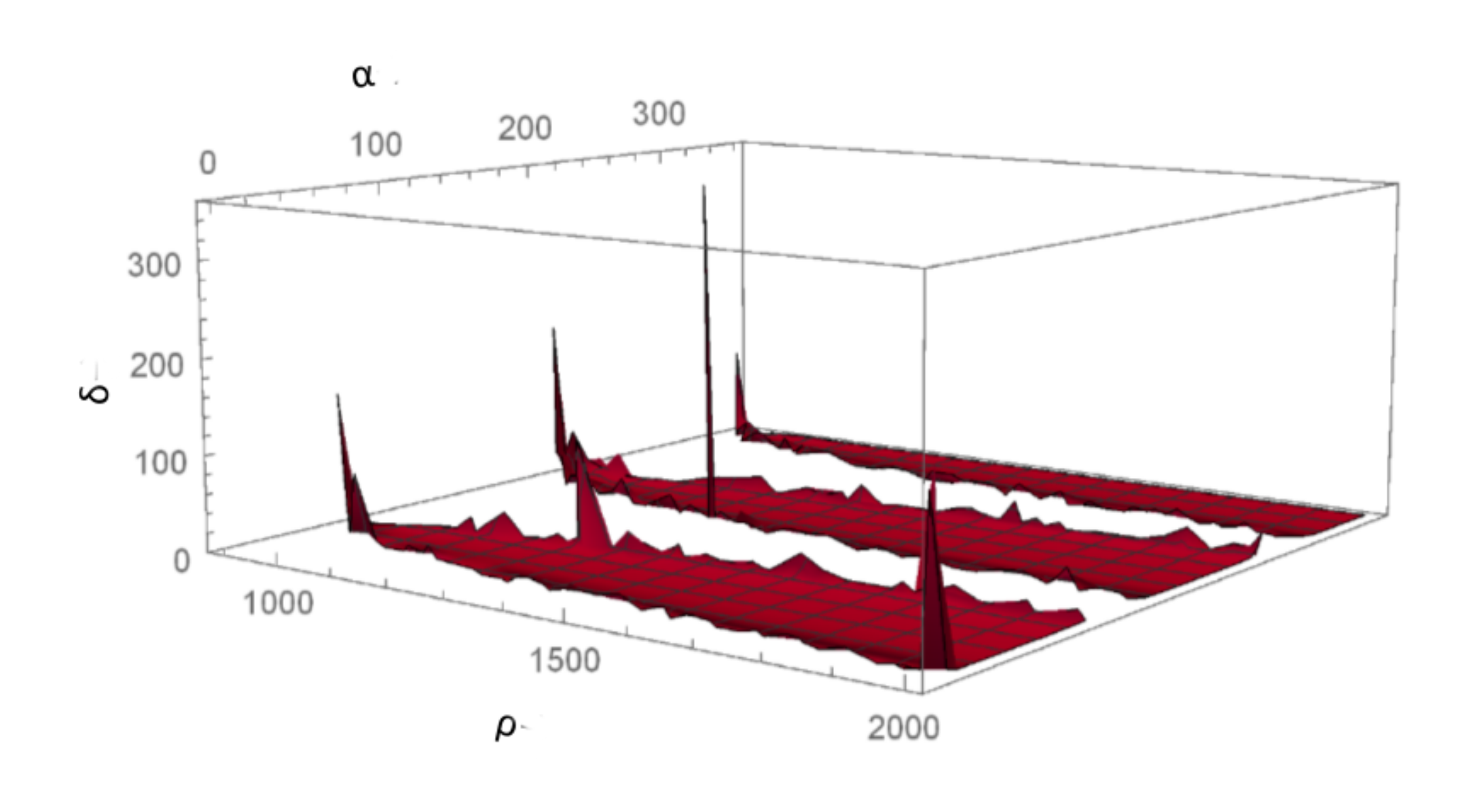}}\end{subfigure}\\
\caption{In {\color{blue}Fig. 3(a)}, the values of $ \delta_{CP} $ within its $ 3\sigma $ bounds phase for different regions in $ \alpha $ space for $ \frac{M^{'}}{M} =10^{-2}$ are shown. {\color{blue}Fig. 3(b)} shows the values of $ \delta_{CP} $ within its $ 3\sigma $ bounds as indicated by current neutrino oscillation global fit {\color{blue}\cite{pc}} for different regions in $ \alpha $ space for $ \frac{M^{'}}{M} =10^{-3}$.}}. 
\label{fig:1}
\end{figure*}
\end{center}

\begin{center}
\begin{figure*}{\includegraphics[height=	7.7cm,width=8cm]{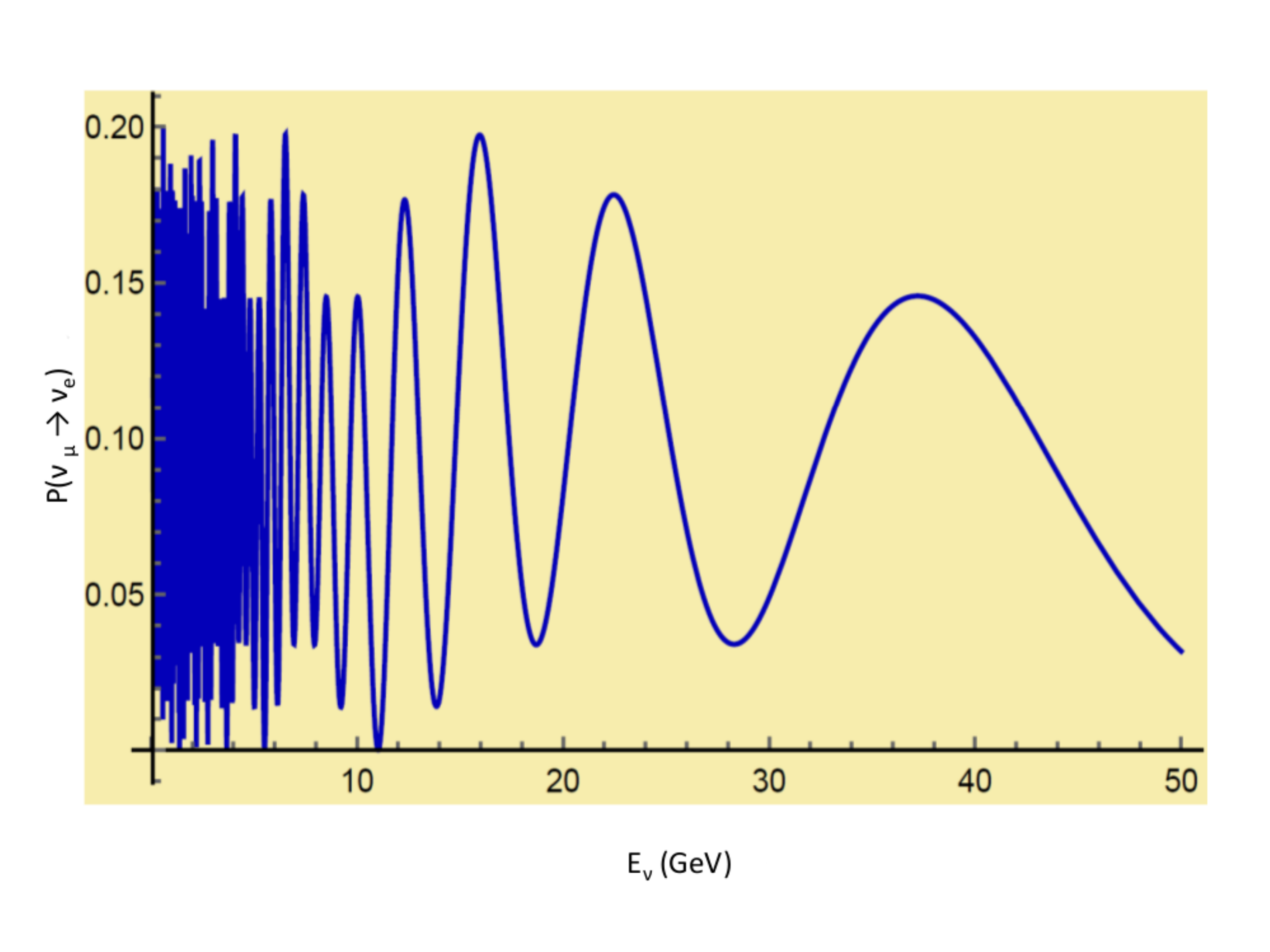}}
\caption{In {\color{blue}Fig. 4} the allowed range of electron neutrino appearance probability at T2K which covers a more restricted region is depicted. The blue region is our model prediction.}
\label{fig:1}
\end{figure*}
\end{center}
\begin{center}
\begin{figure*}[htbp]
\centering{
\begin{subfigure}[]{\includegraphics[height=7.5cm,width=7.6cm]{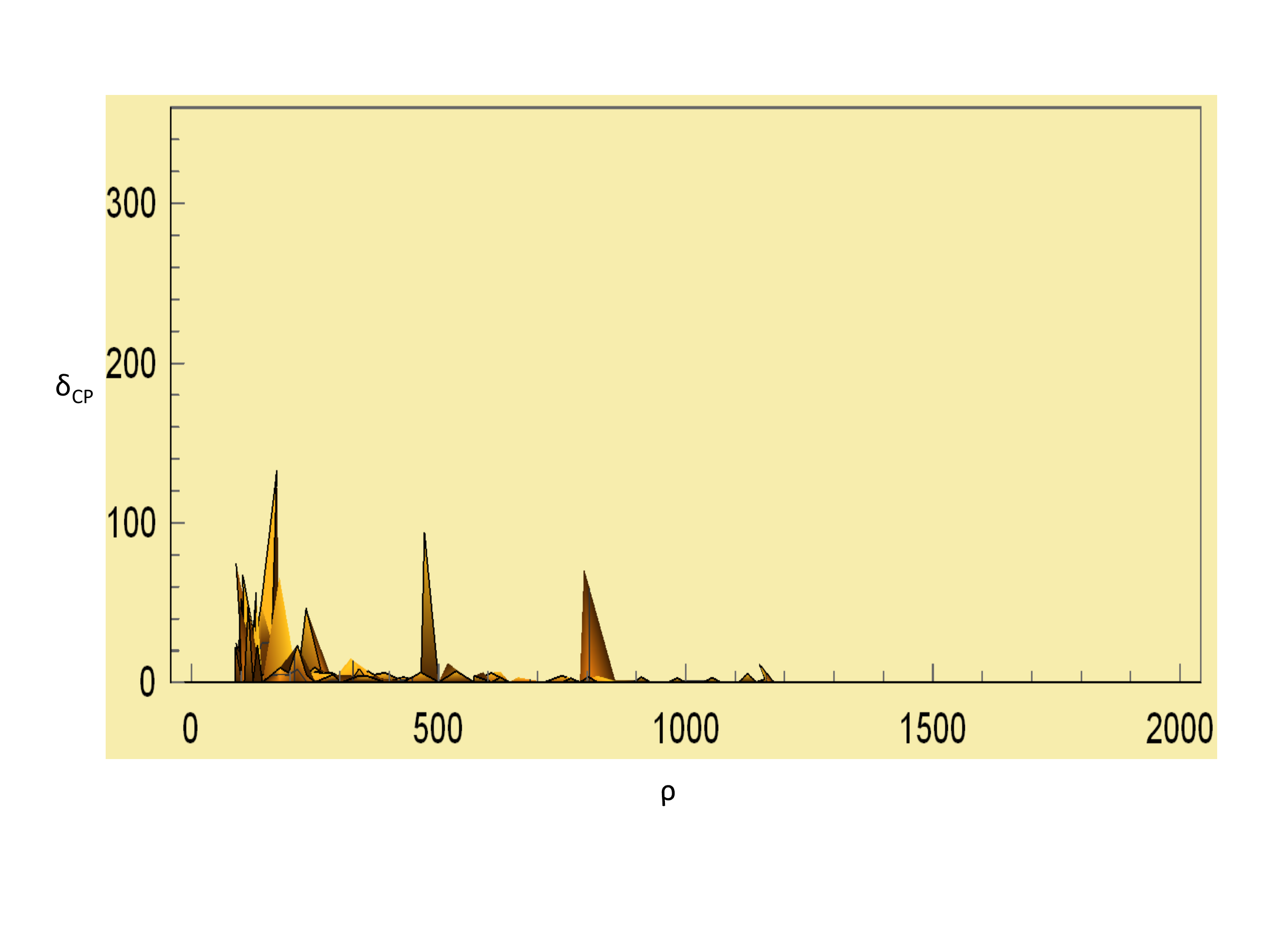}}\end{subfigure}
\begin{subfigure}[]{\includegraphics[height=	7.7cm,width=8cm]{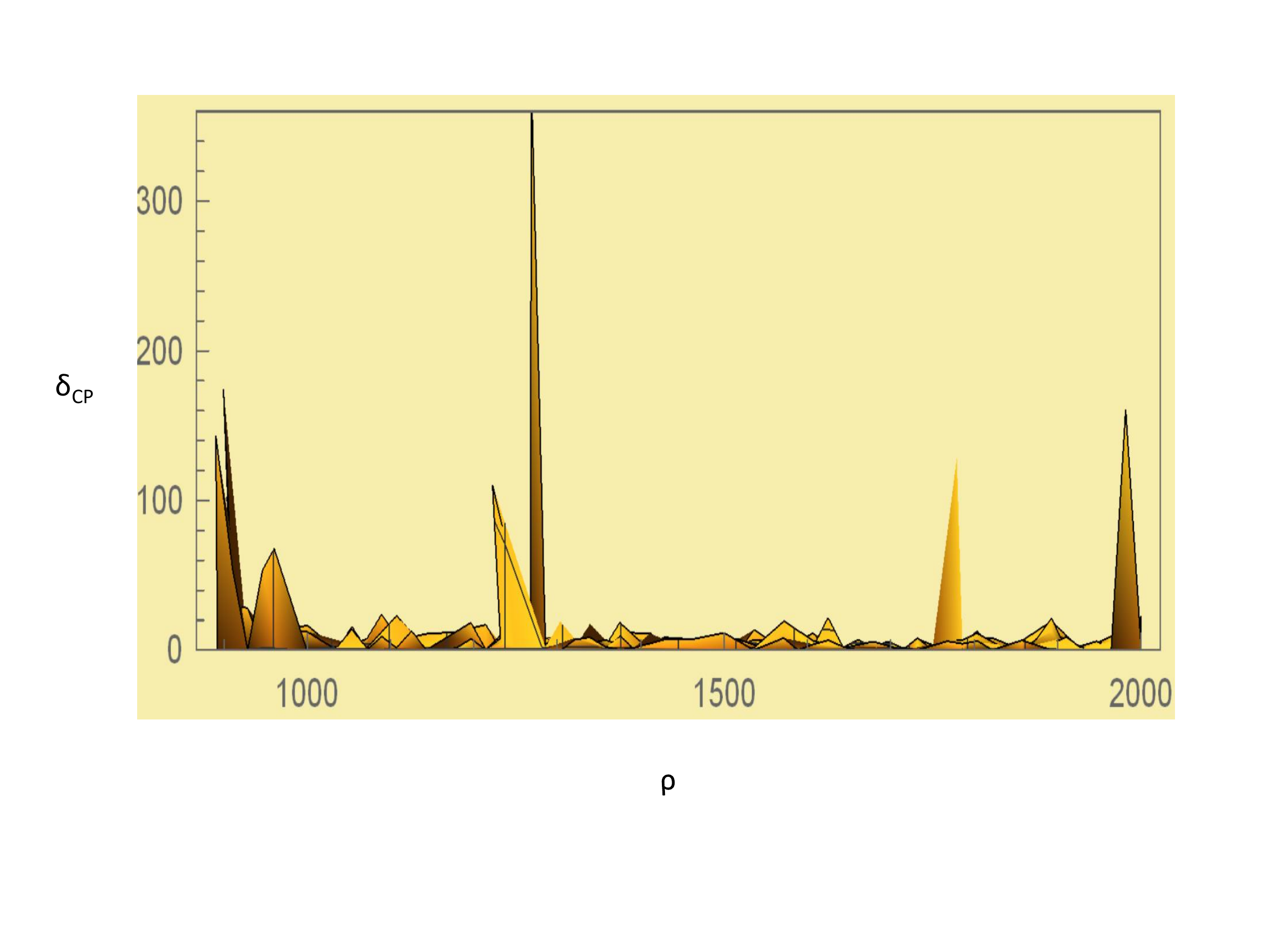}}\end{subfigure}\\
\caption{{\color{blue}Fig. 5(a)} presents the values of $ \delta_{CP} $ within its $ 3\sigma $ bounds phase for different regions in $ \rho$ space for $ \frac{M^{'}}{M} =10^{-2}$. In {\color{blue}Fig. 5(b)} the values of 
$ \delta_{CP} $ within its $ 3\sigma $ bounds as indicated by current neutrino oscillation global fit {\color{blue}\cite{pc}}, for different regions in $ \rho $ space for $ \frac{M^{'}}{M} =10^{-3}$ are illustrated.}}. 
\label{fig:1}
\end{figure*}
\end{center}
The value of $ \delta_{CP} $ depends on the relative predominance between the parameters $ Sin\hspace{0.1cm}\alpha $ and $ \kappa $ or $ \rho$. This dependance is plotted in 
{\color{blue}Figs. 3, 5}. From the right panel in {\color{blue} Fig. 3} the results of our present analysis suggests $ \delta_{CP} $ violation phase to be around $144^{{\circ}}$, corresponding to $ \frac{M^{'}}{M}=10^{-3}$ with $ \alpha\sim 60^{\circ} $. The analysis of No$ \nu $A results shows a preferance for $ \delta_{CP}\sim 0.8 \pi$ suggests our present analysis of $ \delta_{CP} $ phase $ \sim 144^{\circ} $ exactly coincides with the  preferred value. The separate analysis of neutrino and
antineutrino channels can not provide, at present, a sensitive measurement of $ \delta_{CP} $ phase. The CPV phase can therefore be measured by the long-baseline accelerator experiments T2K and NO$ \nu $A, and also by Super-Kamiokande atmospheric neutrino data. Similarly, for $ \frac{M^{'}}{M}=10^{-2}$, we obtain the best fit value of $ \delta_{CP}\sim 0.8 \pi$ in our present analysis corresponding to $ \alpha\sim 310^{\circ} $. 
The predictions made in this analysis can also be tested in currently running and upcoming neutrino oscillation experiments. The predictions made by our model to electron neutrino appearance probability oscillation experiments are displayed in 
{\color{blue} Fig. 4}. This conjecture is for the T2K setup, neglecting matter effects, as an approximation. Clearly, the allowed range of electron neutrino appearance probability at T2K is significantly constrained with respect to the generic
expectation. One finds from the right panel in {\color{blue} Fig. 5}, that
NO$ \nu $A  preferance of $ \delta_{CP}\sim 0.8\pi$ propounds the parameter of perturbation in neutrino sector $ \frac{1}{\rho} $ to be around 5$ \times 10^{-4} $ and 2$ \times 10^{-3} $. From the left panel it is found that preferance of $ \delta_{CP}\sim 0.8\pi$ constrains the parameter of perturbation in neutrino sector $ \frac{1}{\rho} $ to limit itself around 4$ \times 10^{-43} $. The inclusion of reactor data can help to improve the determination of $ \delta_{CP} $ phase, owing to the existing correlation between the CP phase and $ \theta_{13} $. From the results presented in this work, we obtain the best fit value for the CP phase at $ \delta_{CP}\sim 0.8\pi$ for NO. The 
CP conserving value $ \delta $ = 0 is favored with $ \frac{1}{\rho} $ in the region, $ 1\times10^{-3} $ to  $ 5\times10^{-4} $ in NO as can be seen from the left panel in {\color{blue} Fig. 6}. 
\begin{center}
\begin{figure*}[htbp]
\centering{
\begin{subfigure}[]{\includegraphics[height=6.7cm,width=8cm]{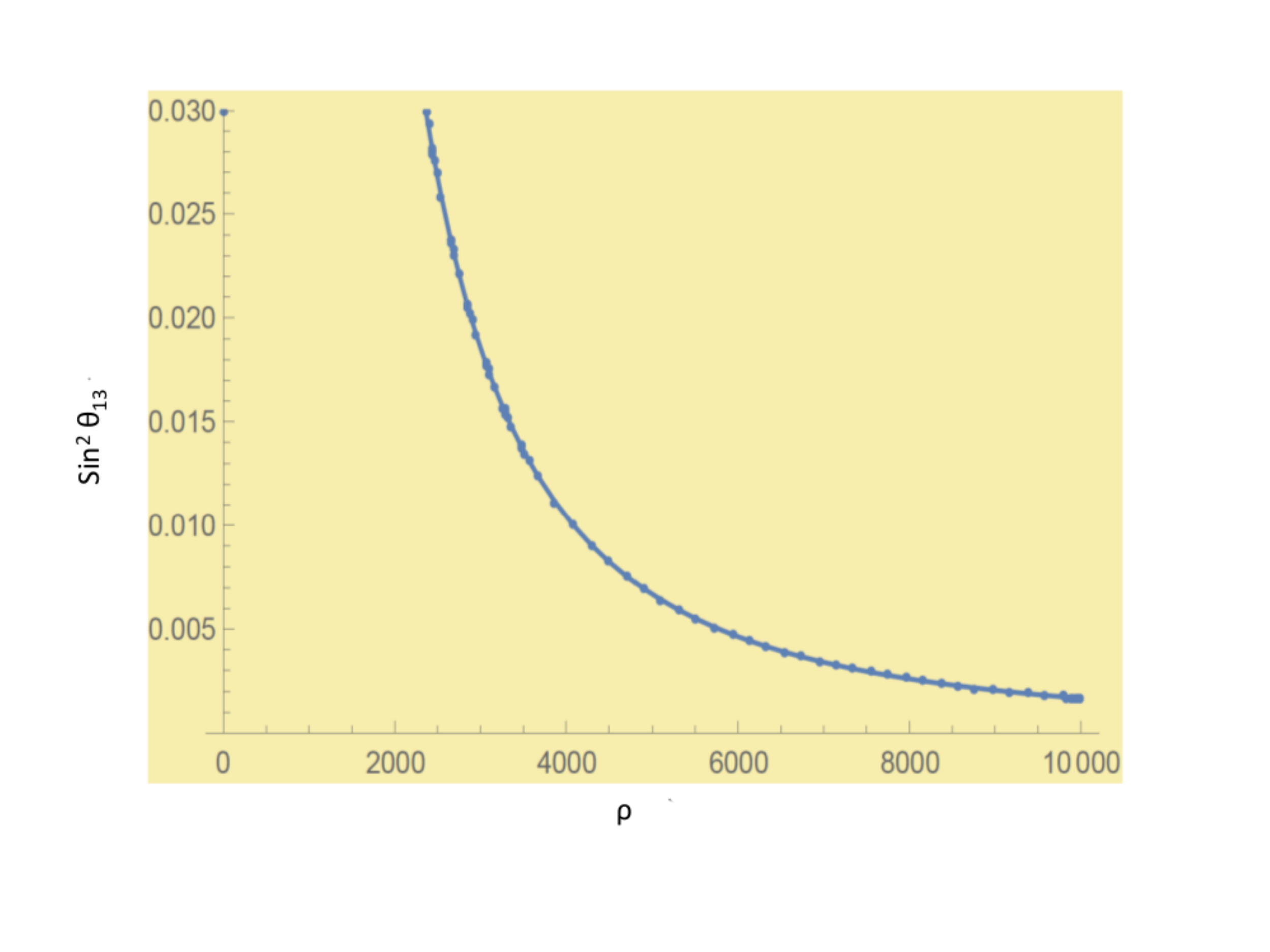}}\end{subfigure}
\begin{subfigure}[]{\includegraphics[height=	6.7cm,width=8.8cm]{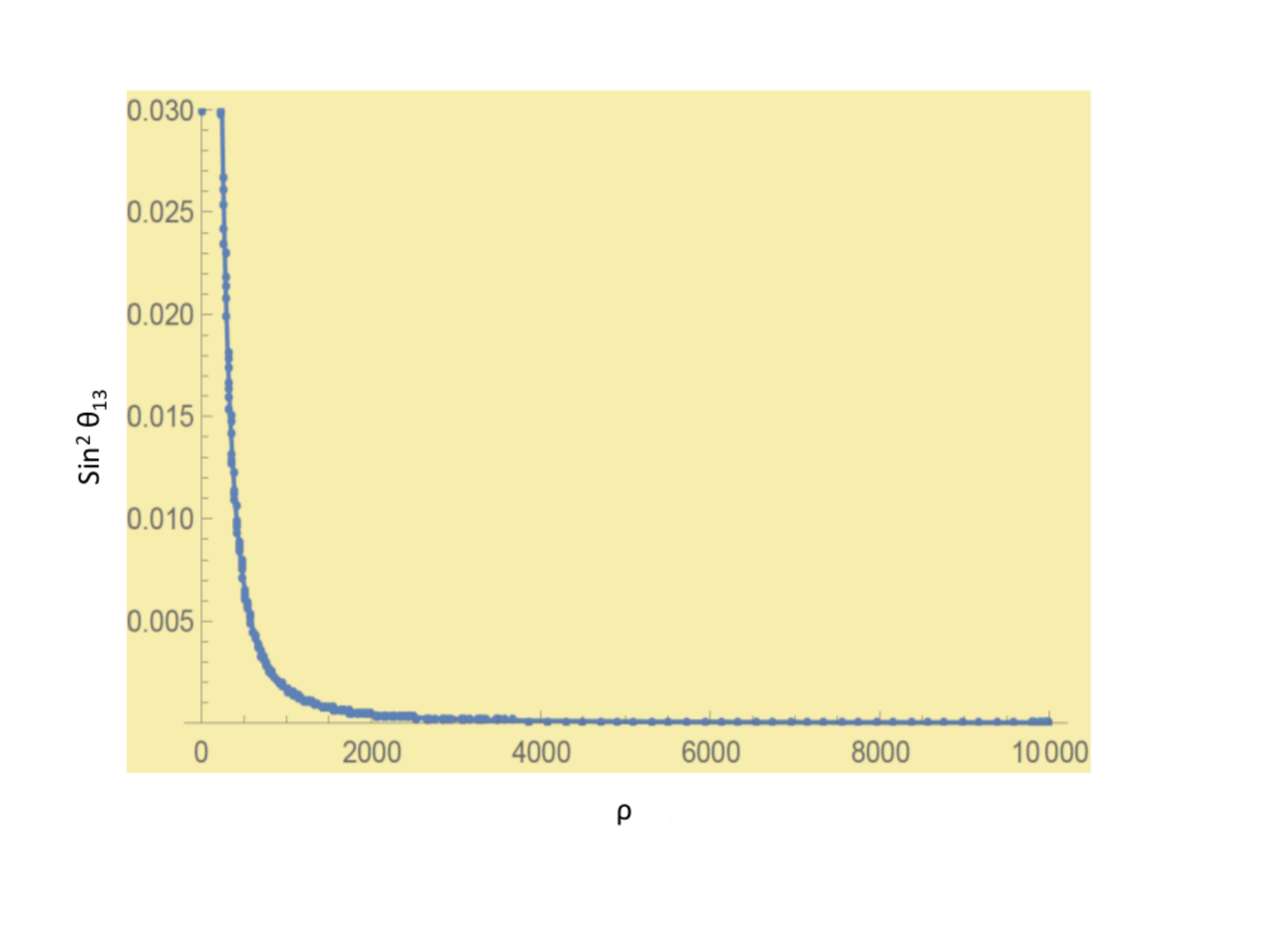}}\end{subfigure}\\

\caption{The plot of sine squared values of mixing angles for maximal $ \delta_{CP} $
through a $ Z_{2} \times Z_{2} $invariant perturbations in neutrino sector are presented in {\color{blue}Fig. 6(a)} and {\color{blue}Fig. 6(b)} which corresponds to $ \frac{M^{'}}{M} =10^{-2}$ and $ \frac{M^{'}}{M} =10^{-3}$ respectively.}} 
\label{fig:1}
\end{figure*}
\end{center}
\begin{center}
\begin{figure*}[htbp]
\includegraphics[height=7cm,width=10cm]{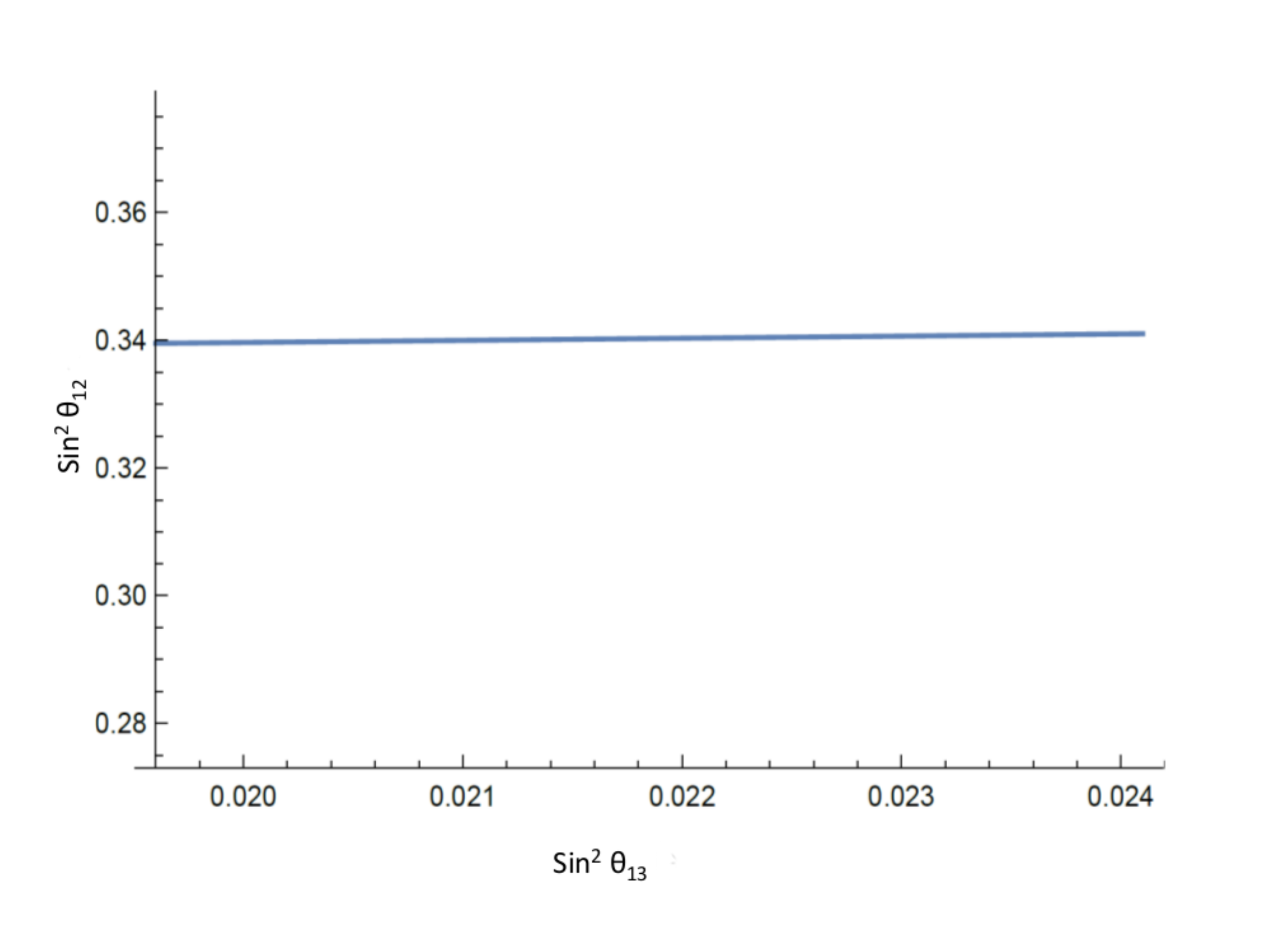}

\caption{In {\color{blue}Fig. 7} the plot of sine squared values of mixing angles for maximal $ \delta_{CP} $ through a $ Z_{2} \times Z_{2} $invariant perturbations in neutrino sector is presented.}
\label{fig:1}
\end{figure*}
\end{center}

\begin{center}
\begin{figure*}[htbp]
\centering{
\begin{subfigure}[]{\includegraphics[height=7.4cm,width=8cm]{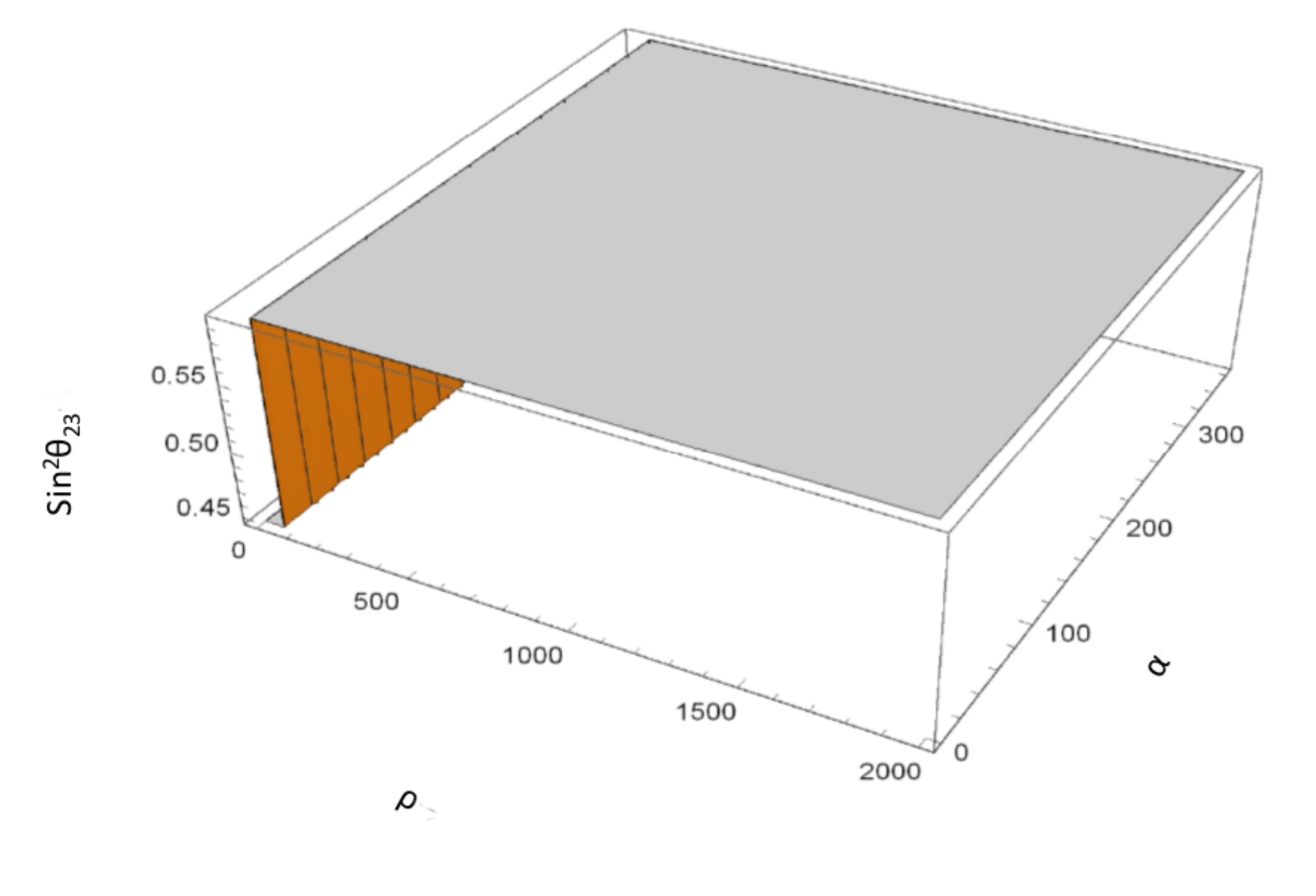}}\end{subfigure}
\begin{subfigure}[]{\includegraphics[height=	7.4cm,width=8cm]{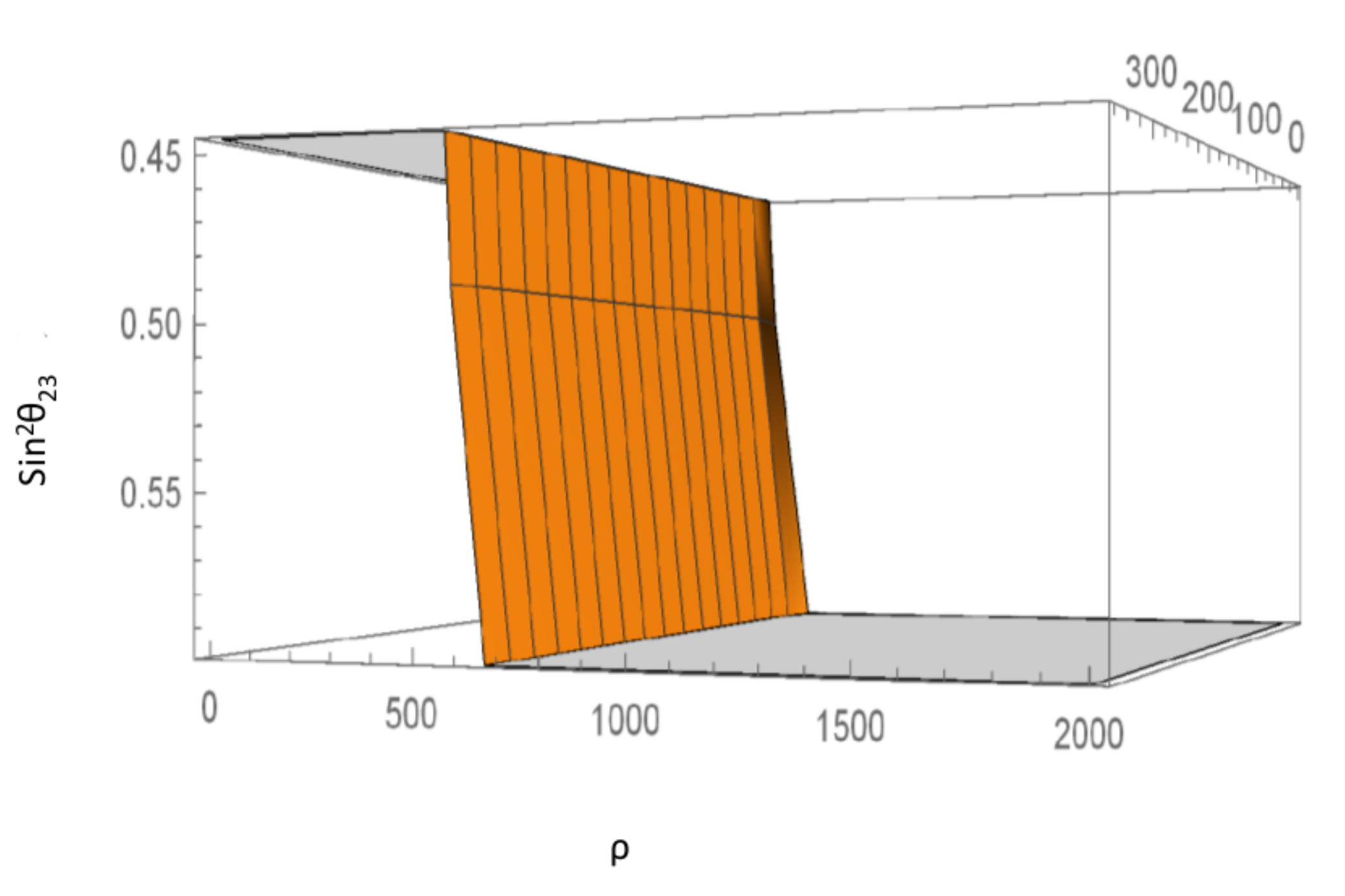}}\end{subfigure}\\

\caption{ {\color{blue}Fig. 8(a)} demonstrates the region in $\rho-\alpha $ space which is consistent with the 3$ \sigma $ constraints on $Sin^{2}\theta_{23}$ for $ \frac{M^{'}}{M} =10^{-2}$. {\color{blue}Fig. 8(b)} illustrates the region in $Sin^{2}\theta_{23}-\rho$ space corresponding to the 3$ \sigma $ bounds on $Sin^{2}\theta_{23}$ for $ \frac{M^{'}}{M}=10^{-3}$.}}. 
\label{fig:1}
\end{figure*}
\end{center}
\begin{center}
\begin{figure*}[htbp]
\includegraphics[height=8cm,width=13cm]{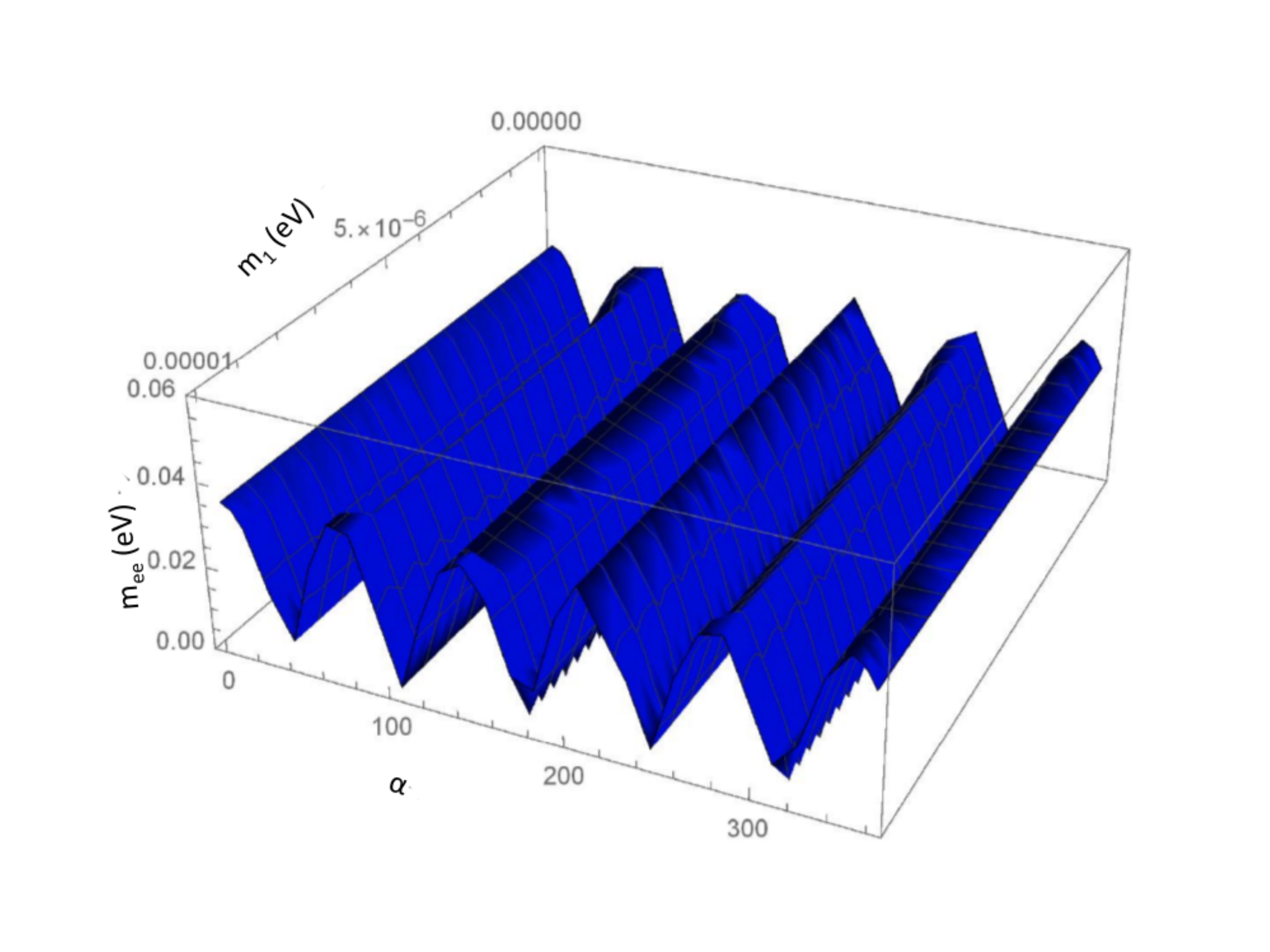}

\caption{$ |m_{ee}|$ prediction for lightest neutrino mass $ m_{1} $ (eV) and $ \alpha $ space in our model.}

\label{fig:1}
\end{figure*}
\end{center}

\begin{center}
\begin{figure*}[htbp]
\includegraphics[height=8cm,width=13cm]{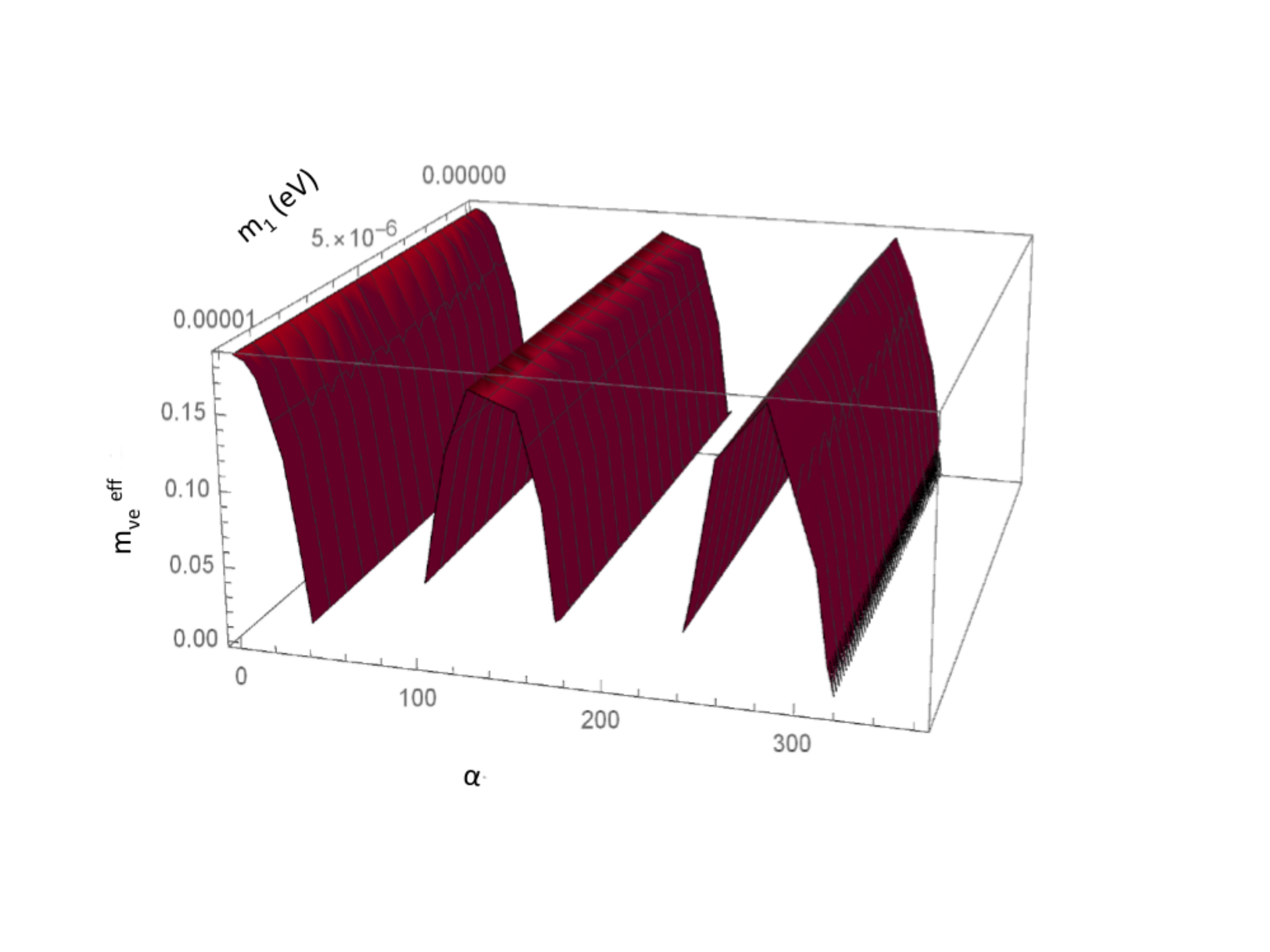}

\caption{$ |m_{\nu_{e}}^{eff}|$ prediction for lightest neutrino mass $ m_{1} $ (eV) and $ \alpha $ space in our model.}
\label{fig:1}
\end{figure*}
\end{center}
Next, we discuss the perturbations in determining the atmospheric mixing angle, $Sin^{2}\theta_{23}$. Accelerator and atmospheric oscillation experiments estimate the disappearance of muon (anti)neutrinos and are mostly sensitive to $Sin^{2}2\theta_{23}$. Thus, one cannot resolve the octant {\color{blue}\cite{GGG,GG1,GG2,GG3}} of the angle. In other words, one cannot decide whether $Sin^{2}2\theta_{23}>0.5$ or $Sin^{2}2\theta_{23}<0.5$. Nonetheless, on account of matter effects in the neutrino trajectories inside the Earth, this degeneracy is slightly broken for atmospheric neutrino oscillation experiment. The quantity $Sin^{2}\theta_{23}$ finds itself in the expressions for appearance channels of these probability experiments. Examining the data from long-baseline accelerators, one finds two essentially degenerate solutions for $Sin^{2}2\theta_{23}$ for both mass orderings, The best fit is obtained for $Sin^{2}\theta_{23} = 0.46$, and a local minimum appears at $Sin^{2}2\theta_{23}$ = 0.57 with $ \Delta\chi^{2} =0.3(0.7)$ for normal mass (inverted mass) ordering. 
In the present analysis the best fit experimental value of $Sin^{2}\theta_{23} = 0.57$  is $ 8.59 \% $ larger than the TBM value of 0.5. To procure this deviation, one needs $ \frac{1}{\rho} $ to be 0.0111 and $ \frac{1}{\rho} $ to be $ 1.54\times10^{-3} $ for $ \frac{M^{'}}{M}=10^{-2}$ and $ \frac{M^{'}}{M}=10^{-3}$ respectively as seen from {\color{blue}Fig. 8}. The above values of $ \frac{1}{\rho} $ lead to small values of $ \kappa $ and hence of $ \delta_{CP} $ phase. Thus there is a strain in obtaining large values of $ \delta_{CP} $ phase and best fit experimental value of $Sin^{2}\theta_{23} $. For $ \kappa> 2 Sin\hspace{0.1cm}\alpha $ we have large CPV phase, and that keeps the value 
of $Sin^{2}\theta_{23} $ close to the TBM value of 0.5  as is evident 
from {\color{blue}Eq. (35)}.
\par 
Owing to the constrained limited nature of the mixing angles  and perturbing factors like $ \frac{1}{\rho} $  and $ \alpha $ in our model, one also gets predictions for $ |m_{ee} |$ and $ |m_{\nu_{e}}^{eff}|$, as shown in {\color{blue}Fig. 9} and {\color{blue}Fig. 10}
 respectively.
\begin{center}
\begin{figure*}[htbp]
\centering{
\begin{subfigure}[]{\includegraphics[height=7.7cm,width=8cm]{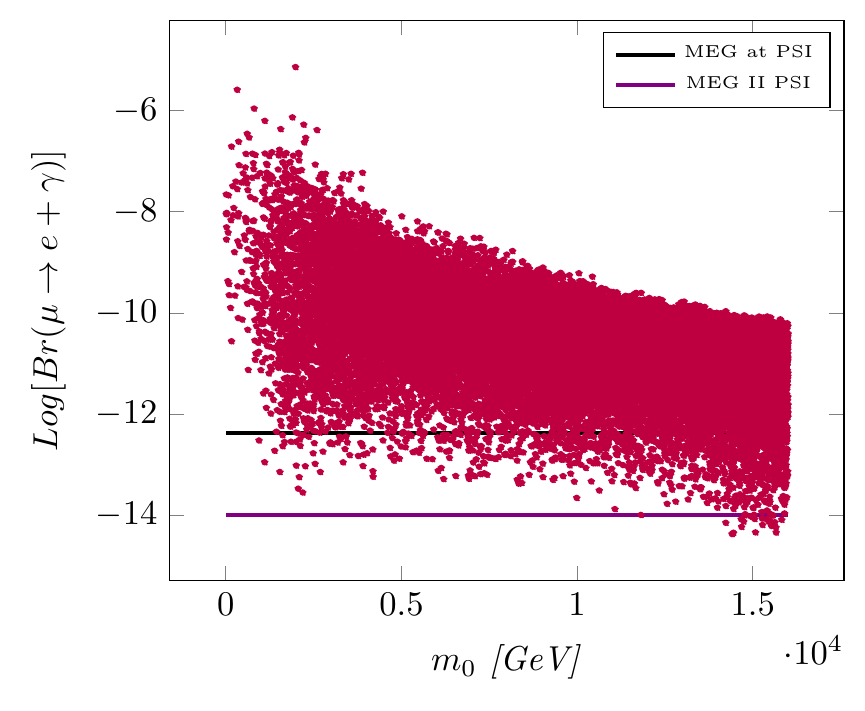}}\end{subfigure}
\begin{subfigure}[]{\includegraphics[height=	7.7cm,width=8cm]{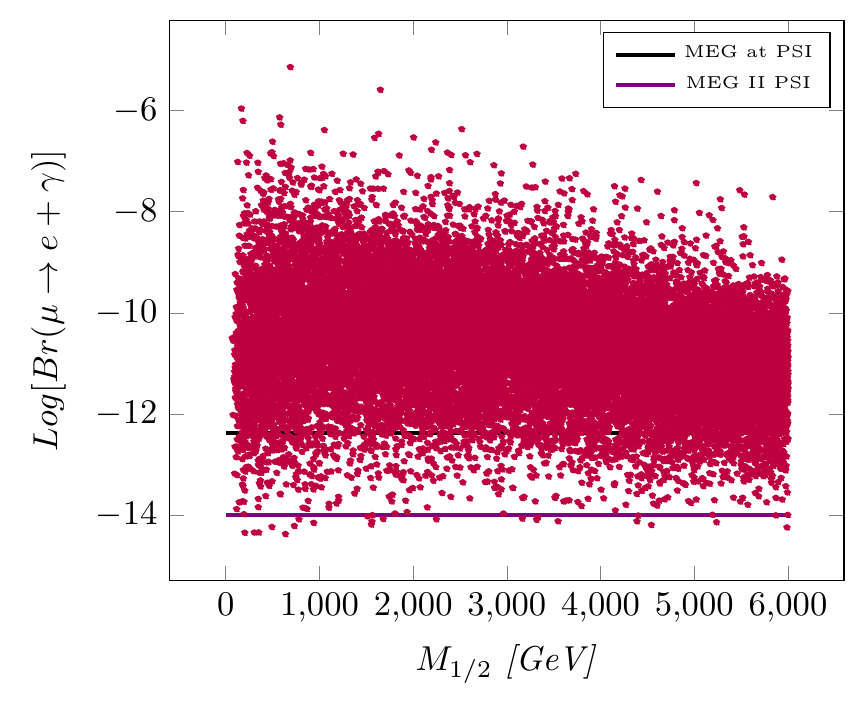}}\end{subfigure}\\
\caption{In {\color{blue}Fig. 11a, 11b}, different horizontal lines black and violet represent the present MEG bound at PSI and future MEG II PSI bounds for BR($ \mu $ $ \rightarrow $ e+ $ \gamma $) respectively.}}. 
\label{fig:1}
\end{figure*}
\end{center}
To test various CLFV processes, like $ \mu  \rightarrow e  \gamma $, $ \tau  \rightarrow e  \gamma $, $ \tau  \rightarrow \mu  \gamma $, the parameters we use in our model are scalar masses $m_{0}$, triliniear coupling, $A_{0}$ and unified gaugino mass
 $ M_{1/2} $. There is also the Higgs potential parameter $ \mu $ and the undetermined ratio of the Higgs VEVs, Tan$\beta$. The present MEG bound at PSI i.e, $ <4.2 \times 10^{-13} $ together with a non zero values of $\theta_{13}$ {\color{blue}\cite{pc}} puts notable constraints on SUSY parameter space. As can be seen from {\color{blue}Fig. 11a}, only small part of the heavy $ m_{0} $ space around 15000 GeV survives for Tan $ \beta $ = 10 in our model restricted by future MEG II bound at PSI  for BR($ \mu \rightarrow e \gamma $) $ \sim 10^{-14} $. {\color{blue}Fig. 11b} reveals that the parameter space $M_{1/2} \geq $ 1 TeV is allowed by present MEG PSI bounds on BR($ \mu \rightarrow e \gamma $), while future MEG limit BR($ \mu \rightarrow e \gamma $) $ \sim 10^{-14} $ excludes almost whole of $M_{1/2}$ space. Very few points around $ \sim $ 1 $ - $ 2 TeV are favourable. In {\color{blue}Fig. 12a, 12b}, we show the correlation among the branching ratios of BR($ \tau  \rightarrow  \mu +  \gamma $) versus  BR($ \mu $ $ \rightarrow $ e+ $ \gamma $) and BR($ \tau  \rightarrow  e +  \gamma $) versus  BR($ \mu $ $ \rightarrow $ e+ $ \gamma $) respectively.
\begin{center}
\begin{figure*}[htbp]
\centering{
\begin{subfigure}[]{\includegraphics[height=7.7cm,width=8cm]{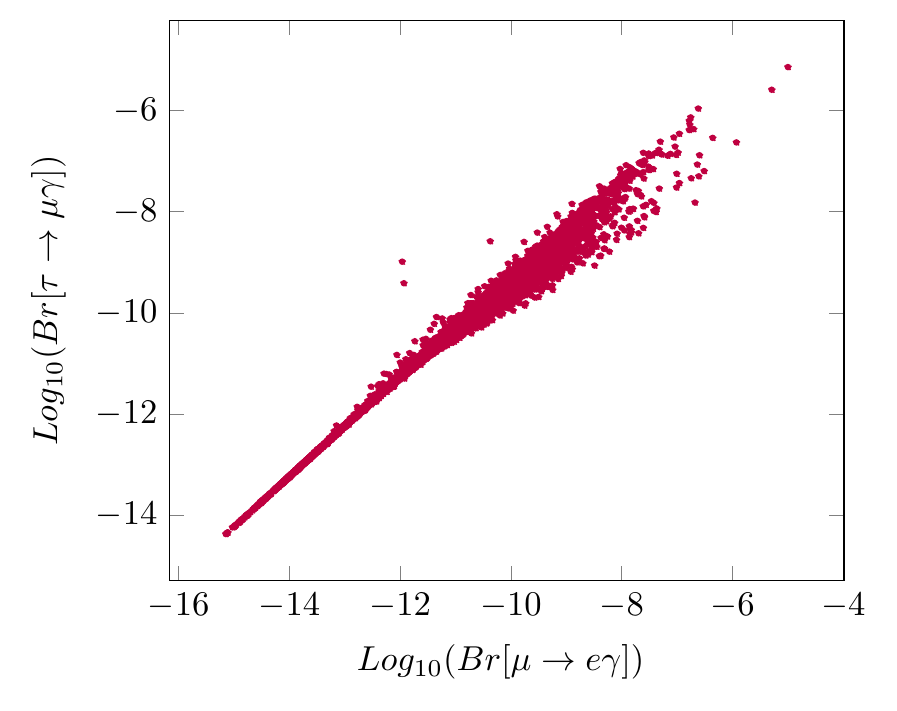}}\end{subfigure}
\begin{subfigure}[]{\includegraphics[height=	7.7cm,width=8cm]{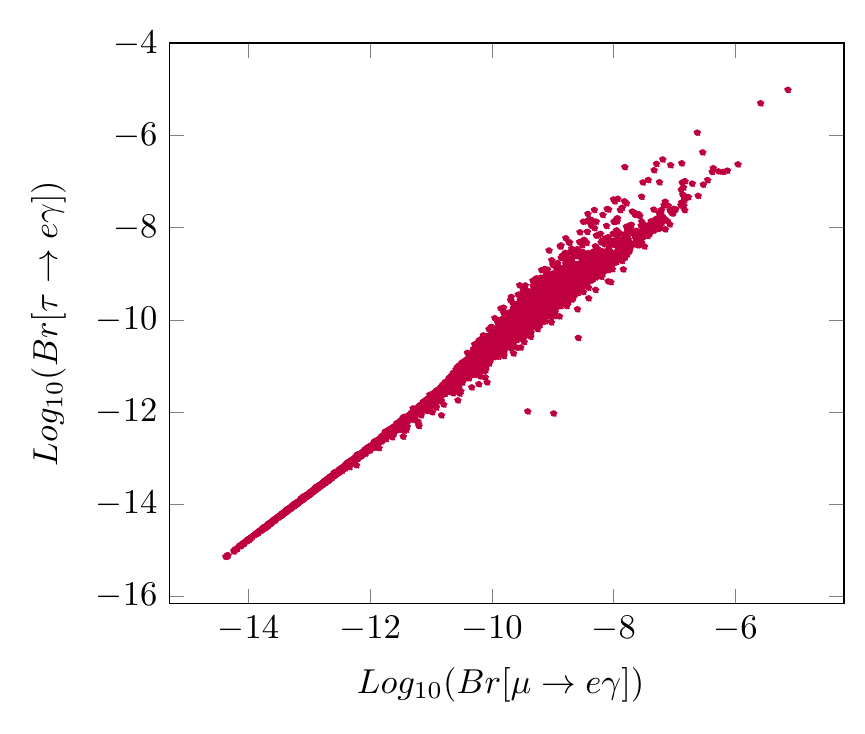}}\end{subfigure}\\
\caption{In {\color{blue}Fig. 12a, 12b} correlations between different LFV decays,  BR($ \tau  \rightarrow  \mu +  \gamma $) versus  BR($ \mu $ $ \rightarrow $ e+ $ \gamma $) and BR($ \tau  \rightarrow  e +  \gamma $) versus  BR($ \mu $ $ \rightarrow $ e+ $ \gamma $) are shown respectively.}}. 
\label{fig:1}
\end{figure*}
\end{center}

\begin{center}
\begin{figure*}[htbp]
\centering{
\begin{subfigure}[]{\includegraphics[height=7.7cm,width=8cm]{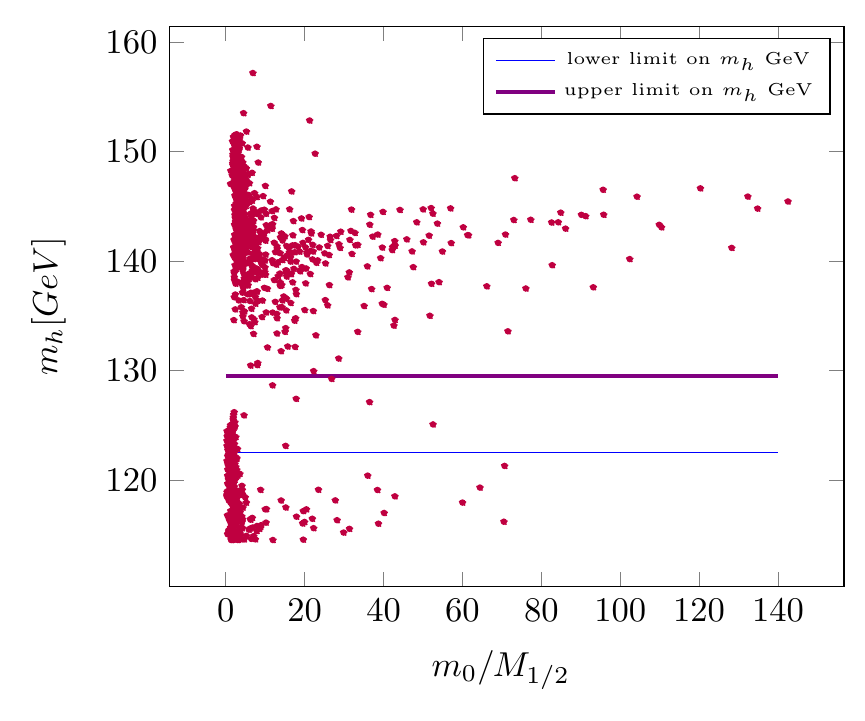}}\end{subfigure}
\begin{subfigure}[]{\includegraphics[height=	7.7cm,width=8cm]{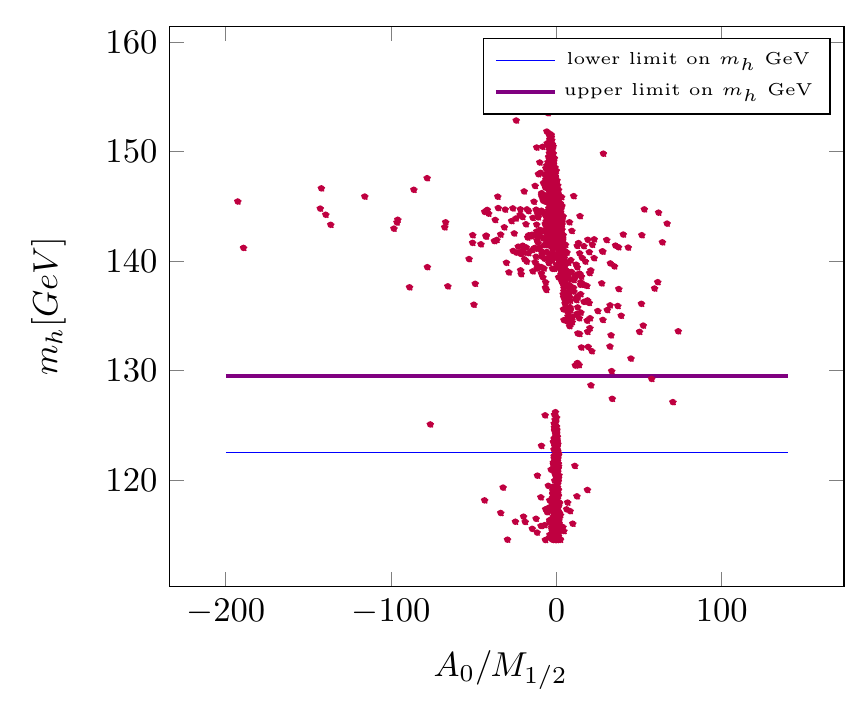}}\end{subfigure}\\
\caption{In {\color{blue}Figs. 13a, 13b} we have shown feasible SUSY parameters space allowed by present bound at MEG at PSI. Different horizontal lines denote the range of Higgs mass as given by the data at LHC, i.e 122.5 $\text{GeV} \le m_{h} \le$ 129.5 $\text{GeV}$ {\color{blue}\cite{GG}}.}}. 
\label{fig:1}
\end{figure*}
\end{center}
The current upper limit on BR($ \mu $ $ \rightarrow $ e+ $ \gamma $)  implies an upper limit BR($ \tau $ $ \rightarrow $ $\mu$+ $ \gamma $) $\sim 10^{-13} $ which is notably smaller than the sensitivity of current generation experiments. Thus, any signatures of CLFV decay $ \tau $ $ \rightarrow $ $\mu$+ $ \gamma $ may rule out the present favoured value of $ \alpha\sim 60^{0} $. In {\color{blue}Fig. 13a, 13b} we plot the lightest Higgs mass $ m_{h} $ as a function of $ m_{0}/M_{1/2} $, $A_{0}/M_{1/2} $ respectively. For the allowed range of Higgs mass as given by the data at LHC, i.e 122.5 $\text{GeV} \le m_{h} \le$ 129.5 $\text{GeV}$, $m_{0}/M_{1/2}$ should be around 5 as allowed by present MEG PSI bounds on BR($ \mu \rightarrow e \gamma $).

\begin{center}
\begin{figure*}[htbp]
\centering{
\begin{subfigure}[]{\includegraphics[height=7.7cm,width=8cm]{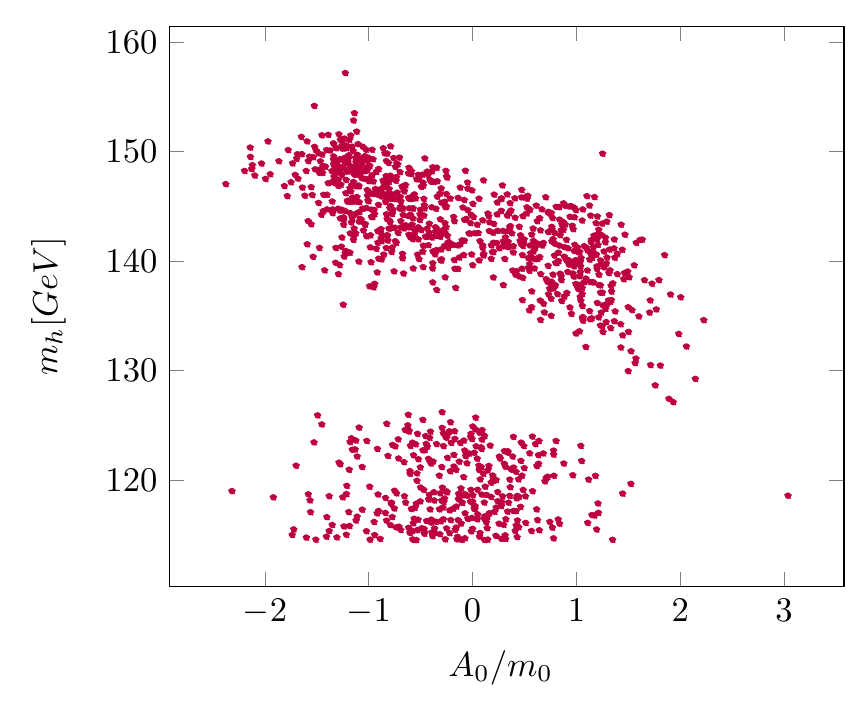}}\end{subfigure}
\begin{subfigure}[]{\includegraphics[height=	7.7cm,width=8cm]{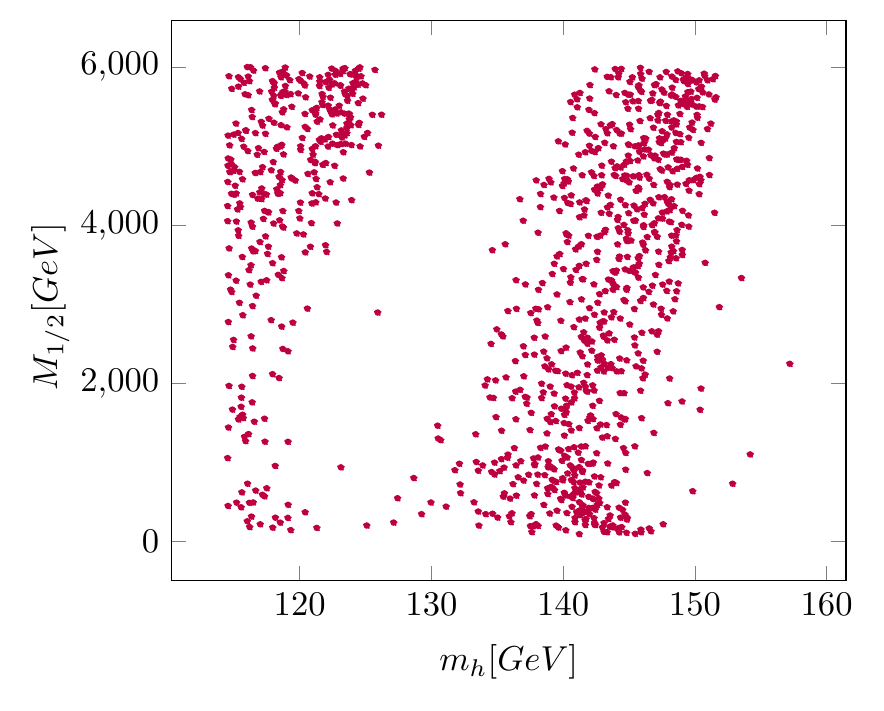}}\end{subfigure}\\
\caption{In {\color{blue}Figs. 14a, 14b} we have shown feasible SUSY parameters space allowed by present bound at MEG at PSI. Different horizontal lines denote the range of Higgs mass as given by the data at LHC, i.e 122.5 $\text{GeV} \le m_{h} \le$ 129.5 $\text{GeV}$ {\color{blue}\cite{GG}}.}}. 
\label{fig:1}
\end{figure*}
\end{center}

\begin{center}
\begin{figure*}[htbp]
\centering{
\begin{subfigure}[]{\includegraphics[height=7.7cm,width=8cm]{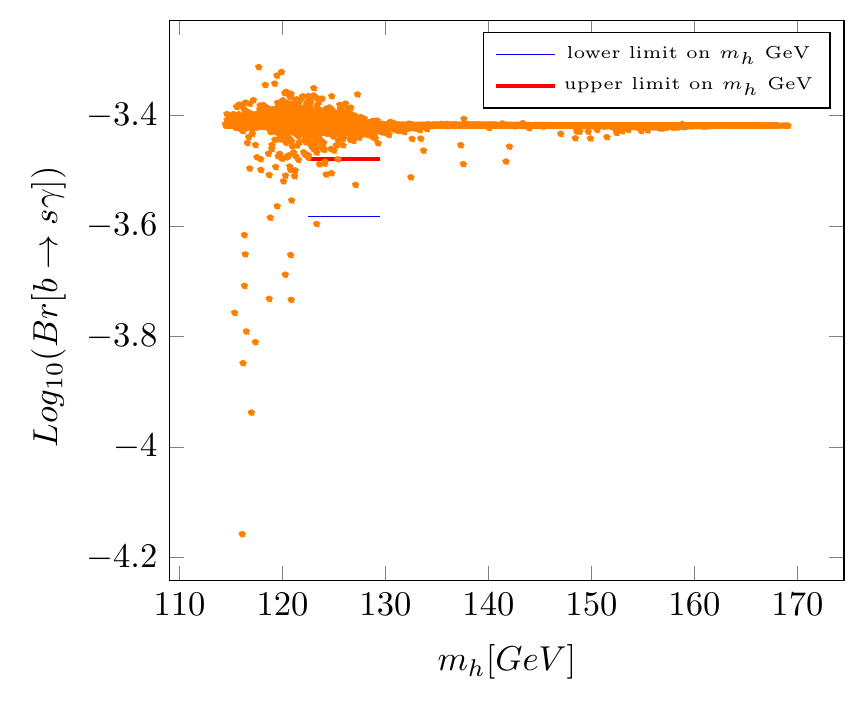}}\end{subfigure}
\begin{subfigure}[]{\includegraphics[height=	7.7cm,width=8cm]{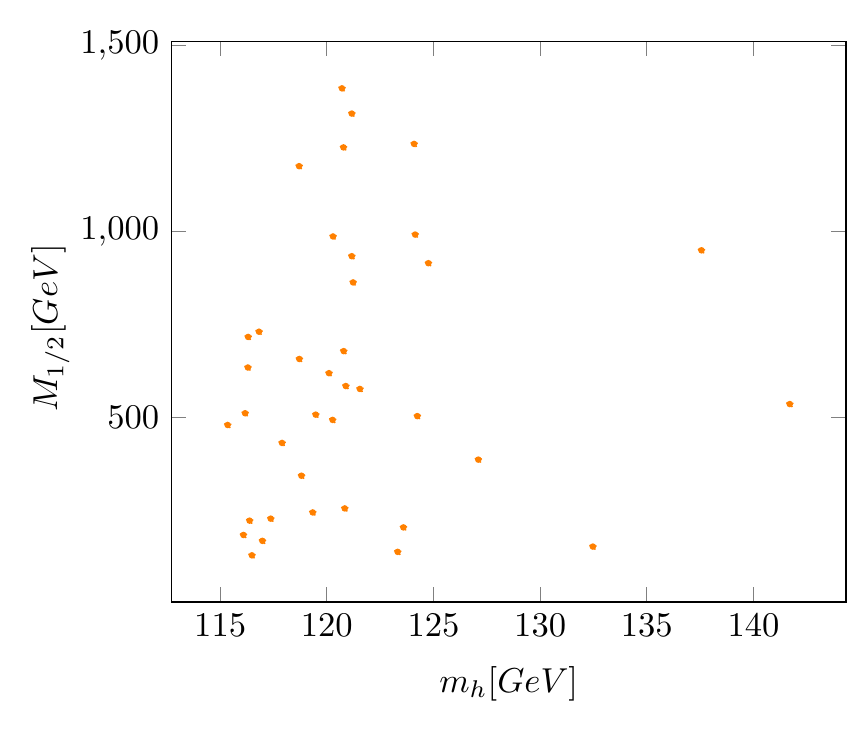}}\end{subfigure}\\
\caption{In {\color{blue}Fig. 15a}, different horizontal lines represent the present  and future bounds for hadronic flavour violation BR(b $ \rightarrow $ s + $ \gamma $). In {\color{blue}Fig. 15b} we have shown SUSY parameters space allowed by present bounds on BR(b $ \rightarrow $ s + $ \gamma $).}}. 
\label{fig:1}
\end{figure*}
\end{center}
In {\color{blue}Fig. 14a}, for the allowed range of Higgs mass, i.e 122.5 $\text{GeV} \le m_{h} \le$ 129.5 $\text{GeV}$, $A_{0}/m_{0}$ should be around -1 to +1. Asymmetry in the value of $A_{0}$, can be seen in {\color{blue}Fig. 14a}. The space $M_{1/2} \geq$ 4 TeV is allowed as can be seen from {\color{blue}Fig. 14b}. In {\color{blue}Fig. 15a} we have presented results for the decay $b \rightarrow s \gamma $. In {\color{blue}Fig. 15b} we have shown SUSY parameters space allowed by present bounds on BR(b $ \rightarrow $ s + $ \gamma $).

\section{Conclusion}
We consider a model based on $ A_{4} $ symmetry which gives the corrections to the TBM form for the leading order neutrino mixing matrix. We present here the phenomenology of a model with $ A_{4} $symmetry which envisage the tribimaximal form for the PMNS matrix. In this model, we have instigated a $ Z_{2} \times Z_{2}$ invariant perturbations in both the charged lepton in the form of $ Sin\hspace{0.1cm}\alpha $ and the neutrino sectors in the form of $ \kappa $. We perceive that perturbations in the neutrino sector leads to allowable values of non zero $ \theta_{13} $ varying within its 3$ \sigma $ range as indicated by current neutrino oscillation global fit {\color{blue}\cite{pc}} and maximal CP violation for $ Sin\hspace{0.1cm}\alpha =0$. The desired value of the CP violating phase $ \delta_{CP} $ lying within its 3$ \sigma $ range can be procured by choosing the fitting and pertinent values for  $ Sin\hspace{0.1cm}\alpha $ term and for the perturbations in neutrino sectors. However, there is a strain in obtaining large values of $ \delta_{CP} $ phase and best fit experimental value of $Sin^{2}\theta_{23} $. For $ \kappa> 2 Sin\hspace{0.1cm}\alpha $ we have large CPV phase, and that keeps the value of $Sin^{2}\theta_{23} $ close to the TBM value of 0.5. Also, our analysis of $ \delta_{CP} $ phase $ \sim 144^{\circ} $ in this model exactly coincides with the  preferred value of $ \delta_{CP}\sim 0.8 \pi$ by the analysis of No$ \nu $A results {\color{blue}\cite{20}}. 
\par 
We have considered leading order corrections in the form of $ Z_{2} \times Z_{2}$ invariant perturbations in neutrino sector after spontaneous breaking of $ A_{4} $ symmetry. The neutrino mixing angles, thus obtained are found to be within the $ 3\sigma $ ranges of their experimental values. The CP violating phase $ \delta_{CP} $ is around $ \sim 144^{\circ} $ in this model. We also studied the variation of the the neutrino oscillation probability $ P(\nu_{\mu}\rightarrow \nu_{e} )$, the effective Majorana mass $ |m_{ee} |$ and $ |m^{eff}_{\nu e} |$ with the lightest neutrino mass $ m_{1} $ in the case of normal hierarchy and found its value to be lower than the experimental upper limit for all allowed values of $ m_{1} \in [0\hspace{0.1cm}eV, 10^{-5} \hspace{0.1cm} eV] $.
\par 
We show that our predicted value of $ \delta_{CP} \sim 144^{0}$ corresponding to $ \frac{M^{'}}{M} = 10^{-3}$ and $ \alpha = 60^{0} $ indicates signatures of various charged LFV channels in a class of $ G_{SM} \times A_{4}\times U(1)_{X}$ model incorporating $ Z_{2} \times Z_{2}$ invariant perturbations in charged lepton and neutrino sector, which is the most interesting feature of our work. These ratios depend on the form of Dirac neutrino yukawa couplings as fixed by Dirac CP phase and on the details of soft SUSY breaking parameters and Tan$\beta$. We have used the  Higgs mass as measured at LHC, non zero reactor mixing angle $ \theta_{13} $ for neutrinos, and latest present and future constraints on BR($ \mu  \rightarrow e  \gamma $). We find that very heavy $ m_{0} $ region is allowed by future MEG bound of BR($ \mu  \rightarrow e  \gamma $) $ \sim 10^{-14} $ and $ M_{1/2} $ values greater than 1 TeV is allowed. We have also shown the predictions for neutrinoless double beta decay in terms of lightest neutrino mass for a mass range of $ m_{1} \in [0\hspace{0.1cm}eV, 10^{-5} \hspace{0.1cm} eV] $, which we epitomize in {\color{blue}Fig. 9, 10}. Ultimately we have estimated the parameters of our model where the branching ratios of $ \mu  \rightarrow e  \gamma $ of the order of $ \sim 10^{-14} $ are calculated, which is well within the latest experimental constraints and is summarised in {\color{blue}Fig. 11}.
\par
To conclude, we have proposed a pragmatic generalization of the TBM ansatz, which in addition to explaining nonzero $ \theta_{13} $, also makes exact and certain
testable predictions for the other parameters of the lepton mixing matrix, including CP violating and CP conserving phases. The $ Z_{2} \times Z_{2}$ invariant perturbations in neutrino sector are characterized in terms of three independent parameters, 
$ \frac{1}{\rho} $, $ \alpha $ and $ \varphi $ which determine all three mixing angles and CP violating phase, leading to several testable predictions. A more comprehensive version of the generalized CP methodology and its potential to produce other hypothetically and realistic ansatz forms for the lepton mixing matrix will be presented in our future work.

\section{Acknowledgement}
GG would like to thank would like to thank University Grants Commission RUSA, MHRD, Government of India for financial support. GG would also like to thank Prof. Probir Roy for useful discussion on this topic. 

\appendix
\section{Basics of $ A_{4} $ group}
$ A_{4} $ is the smallest non Abelian group with an irreducible triplet representation. Alternating group $ A_{4} $ is a group of even permutations of four objects, with a three one dimensional irreducible representation which makes it one of the most favoured group in neutrino mass models.
\par 
$ A_{4} $ group is a non $-$ Abelian, and it is not a direct product of cyclic groups. Group $ A_{4} $ has twelve elements and it is isomorphic to tetrahedral $ T_{d} $.
\par 
$ A_{4} $ group consists of twelve elements which are written in terms of generators of the group $ \textbf{\textit{S}} $ and $T$, the generators satisfy the relation,
\begin{equation}
\textbf{\textit{S}}^{2} = \left( \textbf{\textit{S}}T \right) ^{3} = T^{3} = 1.
\end{equation}
There are three one dimensional irreducible representations of the $ A_{4} $ group defined as 
$$1,\hspace{.5cm} \textbf{\textit{S}}=1,\hspace{.5cm} T=1 $$
$$1^{'},\hspace{.5cm} \textbf{\textit{S}}=1,\hspace{.5cm} T=\omega^{2}$$
$$1^{''},\hspace{.5cm} \textbf{\textit{S}}=1,\hspace{.5cm} T=\omega.$$
The three dimensionals unitary representations of T and S are,
\begin{equation}
T  =  \begin{bmatrix}
1 & 0 & 0\\
0 & \omega^{2}& 0\\
0 & 0 & \omega\\
\end{bmatrix},
\end{equation}
and
\begin{equation}
\textbf{\textit{S}} =  \frac{1}{3}\begin{bmatrix}
-1 & 2 & 2\\
2 & -1 & 2\\
2 & 2 & -1\\
\end{bmatrix}.
\end{equation}
The multiplication rules for the singlet and triplet representations of two generators \textbf{\textit{S}} and T of $ A_{4} $ are, 
$$1\otimes1=1,\hspace{0.3cm}1^{''}\otimes1^{''}=1^{'} $$
$$1^{'}\otimes1^{''}=1,\hspace{0.3cm}1^{'}\otimes1^{'}=1^{''} $$
$$3\otimes3= 3_{1}+3_{2}+1+1^{'}+1^{''} .$$ $ A_{4} $ is a symmetry group of tetrahedron. There are twelve 
independent transformations of the tetrahedron and hence there are twelve group elements as follows:\\
a. four rotations by $120^{0}$ clockwise $\left(\text{as  seen from the vertex}\right)$ which are T-type,\\
b. four rotations by $120^{0}$ anticlockwise $\left(\text{as  seen from the vertex}\right)$,\\
c. three rotations by $180^{0}$ $ - $ \textbf{\textit{S}} type,\\
d. 1 unit operator 1.
\par
$ A_{4} $ has four irreducible representations which are three singlets $ -1,1^{'},1^{''} $ and one triplet. The products of singlets are:
$$1\otimes1=1,\hspace{0.3cm}1^{''}\otimes1^{''}=1^{'} $$
$$1^{'}\otimes1^{''}=1,\hspace{0.3cm}1^{'}\otimes1^{'}=1^{''} .$$
If we consider two triplets
$a=\left( a_{1}, a_{2}, a_{3}\right) $, $b=\left( b_{1}, b_{2}, b_{3}\right) $
then one can write,
\begin{eqnarray}
\left(ab\right)_{1} = a_{1}b_{1} + a_{2}b_{2} + a_{3}b_{3},\\
\left(ab\right)_{1^{'}} = a_{1}b_{1} + \omega^{2}a_{2}b_{2} + \omega a_{3}b_{3},\\
\left(ab\right)_{1^{''}} = a_{1}b_{1} + \omega a_{2}b_{2} + \omega^{2}a_{3}b_{3},\\
\left(ab\right)_{3_{1}} = a_{2}b_{3} + a_{3}b_{1} + a_{1}b_{2},\\
\left(ab\right)_{3_{2}} = a_{3}b_{2} + a_{1}b_{3} + a_{2}b_{1},\\
\omega^{3} =1.
\end{eqnarray}
In the basis of triplet representations
\begin{equation}
\textbf{\textit{S}} =  \begin{bmatrix}
1 & 0 & 0\\
0 & -1 & 0\\
0 & 0 & -1\\
\end{bmatrix},
\end{equation}
\begin{equation}
T =  \begin{bmatrix}
0 & 1 & 0\\
0 & 0 & 1\\
1 & 0 & 0\\
\end{bmatrix}.
\end{equation}
One generates 12 real $3\times3$ matrix group elements, after multiplication of the matrices together in all possible ways, like
\begin{equation}
\textbf{\textit{S}}^{2}=  \frac{1}{9}\begin{bmatrix}
-1 & 2 & 2\\
2& -1 & 2\\
2 & 2 & -1\\
\end{bmatrix}\begin{bmatrix}
-1 & 2 & 2\\
2& -1 & 2\\
2 & 2 & -1\\
\end{bmatrix}
=\frac{1}{9}\begin{bmatrix}
9 & 0 & 0\\
0& 9 & 0\\
0 & 0 & 9\\
\end{bmatrix}
=\begin{bmatrix}
1 & 0 & 0\\
0& 1& 0\\
0 & 0 & 1\\
\end{bmatrix}.
\end{equation}
$ A_{4} $ has four classes {\color{blue}\cite{39}} denoted by,
\begin{equation}
a) \hspace{.2cm} C_{1} :\begin{bmatrix}
1 & 0 & 0\\
0& 1& 0\\
0 & 0 & 1\\
\end{bmatrix},
\end{equation}
which is the $3\times3$ matrix representation of $ A_{4} $ elements in $ C_{1} $. Likewise we have,
\begin{equation}
b) \hspace{.2cm} C_{2} :\begin{bmatrix}
1 & 0 & 0\\
0& -1& 0\\
0 & 0 & -1\\
\end{bmatrix},
\begin{bmatrix}
-1 & 0 & 0\\
0& 1& 0\\
0 & 0 & -1\\
\end{bmatrix},
\begin{bmatrix}
-1 & 0 & 0\\
0& -1& 0\\
0 & 0 & 1\\
\end{bmatrix}.
\end{equation}
\begin{equation}
c) \hspace{.2cm} C_{2} :\begin{bmatrix}
0 & 1 & 0\\
0& 0& 1\\
1 & 0 & 0\\
\end{bmatrix},
\begin{bmatrix}
0 & -1 & 0\\
0& 0& -1\\
1 & 0 & 0\\
\end{bmatrix},
\begin{bmatrix}
0 & -1 & 0\\
0& 0& 1\\
-1 & 0 & 0\\
\end{bmatrix},
\begin{bmatrix}
0 & 1 & 0\\
0& 0& -1\\
-1 & 0 & 0\\
\end{bmatrix}.
\end{equation}
\begin{equation}
d) \hspace{.2cm} C_{2} :\begin{bmatrix}
0 & 0 & 1\\
1& 0& 0\\
0 & 1 & 0\\
\end{bmatrix},
\begin{bmatrix}
0 & 0 & -1\\
1& 0& 0\\
0 & -1 & 0\\
\end{bmatrix},
\begin{bmatrix}
0 & 0 & 1\\
-1& 0& 0\\
0 & -1 & 0\\
\end{bmatrix},
\begin{bmatrix}
0 & 0& -1\\
-1& 0& 0\\
0 & 1 & 0\\
\end{bmatrix},
\end{equation}
where $ Z_{3} $ elements are
$$ \begin{bmatrix}
1 & 0 & 0\\
0& 1& 0\\
0 & 0 & 1\\
\end{bmatrix}, \begin{bmatrix}
0 & 1& 0\\
0& 0& 1\\
1 & 0 & 0\\
\end{bmatrix},\begin{bmatrix}
0 & 0 & 1\\
1& 0& 0\\
0 & 1 & 0\\
\end{bmatrix}$$
and $ Z_{2} \times Z_{2}$ elements in this classification are;
\begin{equation}
\begin{bmatrix}
1 & 0 & 0\\
0& 1& 0\\
0 & 0 & 1\\
\end{bmatrix}, \begin{bmatrix}
1 & 0 & 0\\
0& -1& 0\\
0 & 0 & -1\\\end{bmatrix},\begin{bmatrix}
-1 & 0 & 0\\
0& 1& 0\\
0 & 0 & -1\\
\end{bmatrix},\begin{bmatrix}
-1 & 0 & 0\\
0& -1& 0\\
0 & 0 & 1\\
\end{bmatrix}.
\end{equation}


\begin{thebibliography}{}

\bibitem{GG} Kalpana Bora, Gayatri Ghosh, \textit{Charged Lepton Flavor Violation $\mu \rightarrow e \gamma$ in $ \mu $-$ \tau $ Symmetric SUSY SO(10) mSUGRA, NUHM, NUGM and NUSM Theories and LHC}, {\color{blue}Eur. Phys. J. \textbf{C75}, 9, 428, (2015), arXiv:1410.1265 [hep-ph]}.
\bibitem{a}P.F. de Salas, D.V. Forero, C.A. Ternes, M. Tortola and J.W.F. Valle, Status of neutrino oscillations 2018: 3$ \sigma $ hint for normal mass ordering and improved CP sensitivity, {\color{blue}Phys. Lett. B 782 (2018) 633 [arXiv:1708.01186] [ IN SPIRE ]}.
\bibitem{b} D.V. Forero, M. Tortola and J.W.F. Valle, Neutrino oscillations refitted, {\color{blue}Phys. Rev. D 90 (2014) 093006 [arXiv:1405.7540] [ IN SPIRE ]}.
\bibitem{c} D.V. Forero, M. Tortola and J.W.F. Valle, Global status of neutrino oscillation parameters after Neutrino-2012, {\color{blue}Phys. Rev. D 86 (2012) 073012 [arXiv:1205.4018] [ IN SPIRE ]}.
\bibitem{d}T. Schwetz, M. Tortola and J.W.F. Valle, Where we are on $ \theta_{13} $: addendum to ‘Global neutrino data and recent reactor fluxes: status of three-flavour oscillation parameters’,  {\color{blue}New J. Phys. 13 (2011) 109401 [arXiv:1108.1376] [ IN SPIRE ]}.
\bibitem{1} SNO collaboration, Direct evidence for neutrino flavor transformation from neutral current interactions in the Sudbury Neutrino Observatory, Phys. Rev. Lett. 89 (2002) 011301 [{\color{blue}nucl-ex/0204008}] [ {\color{blue}IN SPIRE }].
\bibitem{2} Super-Kamiokande collaboration, Evidence for oscillation of atmospheric neutrinos, {\color{blue}Phys. Rev. Lett. 81 (1998) 1562 [hep-ex/9807003] [ IN SPIRE ]}.
\bibitem{3} A.B. McDonald, Nobel Lecture: The Sudbury Neutrino Observatory: Observation of flavor change for solar neutrinos, {\color{blue}Rev. Mod. Phys. 88 (2016) 030502}.
\bibitem{4} T. Kajita, Nobel Lecture: Discovery of atmospheric neutrino oscillations, {\color{blue}Rev. Mod. Phys. 88 (2016) 030501}.
\bibitem{5} KamLAND collaboration, First results from KamLAND: Evidence for reactor anti-neutrino disappearance, {\color{blue}Phys. Rev. Lett. 90 (2003) 021802 [hep-ex/0212021] [ IN SPIRE ]}.
\bibitem{39d}  K. Babu, E. Ma, and J. Valle, Underlying A(4) symmetry for the neutrino mass matrix and the quark mixing matrix," {\color{blue}Phys.Lett. B552, 207{213, 2003}.}
\bibitem{39}  E. Ma and G. Rajasekaran, "Softly broken A(4) symmetry for nearly degenerate neutrino masses," {\color{blue}Phys.Rev. , D64, p. 113012, 2001}.
\bibitem{39a} G. Altarelli and F. Feruglio, Tri-bimaximal neutrino mixing from discrete symmetry in extra dimensions," {\color{blue}Nucl.Phys., B720, 64{88}, 2005}.
\bibitem{39b} G. Altarelli and F. Feruglio, Tri-bimaximal neutrino mixing, A(4) and the modular symmetry," {\color{blue}Nucl.Phys., B741, 215{235}, 2006};
\bibitem{39c} X.-G. He, Y.-Y. Keum, and R. R. Volkas, A(4) avor symmetry breaking scheme for understanding quark and neutrino mixing angles," {\color{blue}JHEP, 0604, 039, 2006.}
\bibitem{e} J. Schechter and J. W. F. Valle, {\color{blue} Phys. Rev. D 22, 2227 (1980)}.
\bibitem{6} P. F. Harrison, D. H. Perkins, and W. G. Scott, {\color{blue}Phys. Lett. B
530, 167 (2002),[arXiv: hep-ph/0202074]}.
\bibitem{dhrm} I. Stancu, D. V. Ahluwalia, L/E-Flatness of the Electron-Like Event Ratio in Super-Kamiokande and a Degeneracy in Neutrino Masses, {\color{blue} Phys.Lett.B460, 1999, arXiv:hep-ph/9903408}.
\bibitem{7} F. P. An et al., {\color{blue}Phys. Rev. D 95, 072006 (2017)}.
\bibitem{8} M. Y. Pac (RENO Collaboration), {\color{blue}Proc. Sci., NuFact2017,
38 (2018), https://pos.sissa.it/295/038/pdf}; Y. Abe et al. (Double Chooz Collaboration), {\color{blue}J. High Energy Phys. 10 (2014) 086; 02 (2015) 074}.
\bibitem{GC} Gayatri Ghosh, Significance of Broken $\mu-\tau$ Symmetry in correlating $\delta_{CP}, \theta_{13}$, Lightest Neutrino Mass and Neutrinoless Double Beta Decay, , {\color{blue} Adv.High Energy Phys. 2021 (2021) 9563917}.
\bibitem{BK} Biswajit Karmakar, Arunansu Sil, Spontaneous CP Violation in Lepton-sector a common origin for $\theta_{13}$, Dirac CP phase and leptogenesis, {\color{blue} Phys. Rev. D 93, 013006 (2016), arXiv:1509.07090 [hep-ph]}.
\bibitem{Pcheng1} Hong-Jian He and Fu-Rong Yin, Common Origin of $\mu-\tau$ and CP Breaking in Neutrino Seesaw, Baryon Asymmetry, and Hidden Flavor Symmetry, {\color{blue} Phys. Rev. D84 (2011) 033009 [arXiv:1104.2654]}.
\bibitem{Pcheng2} S.F. Ge, H.J. He, and F.R. Yin, Common Origin of Soft $\mu-\tau$  and CP Breaking in Neutrino Seesaw and the Origin of Matter, {\color{blue}JCAP 1005 (2010) 017 [arXiv:1001.0940]}
\bibitem{Pcheng3} Hong-Jian He and Xunjie Xu, Octahedral Symmetry with Geometrical Breaking: New Prediction for Neutrino Mixing Angle $\theta_{13}$ and CP 
Violation,  {\color{blue} Phys. Rev. D86 (2012) 111301 (R) [arXiv:1203.2908].}
\bibitem{Pcheng4} Hong-Jian He, Werner Rodejohann, Xun-Jie Xu, Origin of Constrained Maximal CP Violation in Flavor Symmetry, {\color{blue} Phys. Lett. B751 (2015) 586 [arXiv:1507.03541].}
\bibitem{9} RENO collaboration, Measurement of Reactor Antineutrino Oscillation Amplitude and Frequency at RENO,{\color{blue} Phys. Rev. Lett. 121 (2018) 201801 [arXiv:1806.00248] [ INSPIRE ].}
\bibitem{10}Daya Bay collaboration, Measurement of the Electron Antineutrino Oscillation with 1958 Days of Operation at Daya Bay, {\color{blue}Phys. Rev. Lett. 121 (2018) 241805 [arXiv:1809.02261] [ IN SPIRE ]}.

\bibitem{pa}Moinul Hossain Rahat, Pierre Ramond, and Bin Xu, Asymmetric tribimaximal texture {\color{blue}Phys. Rev. D 98, 055030, arXiv:1805.10684 [hep-ph] [ INSPIRE ]}.

\bibitem{ka} M. Jay Pérez, Moinul Hossain Rahat, Pierre Ramond, Alexander J. Stuart, and Bin Xu, {\color{blue} Phys. Rev. D 100, 075008, arXiv:1907.10698 [hep-ph] [ INSPIRE ]}.  
\bibitem{si} Moinul Hossain Rahat, Leptogenesis from the asymmetric texture, {\color{blue}Phys. Rev. D 103, 035011, arXiv:2008.04204[hep-ph]}. 
\bibitem{ha} M.Jay Perez, Moinul Hossain Rahat, Pierre Ramond, Alexander J.Stuart and Bin Xu, Tribimaximal mixing in the $ SU(5) \times  \tau_{13}$ texture, {\color{blue} Phys. Rev.D\textbf{101}, 075018}; Chee Sheng Fong, Moinul Hossain Rahat, Shaikh Saad, Low-scale Resonant Leptogenesis in SU(5) GUT with $\tau_{13}$ Family Symmetry, {\color{blue}arXiv:2103.14691 [hep-ph].}



\bibitem{11} G.C. Branco, R. Gonzalez Felipe, M.N. Rebelo, H. Serodio {\color{blue} Phys.Rev. D79 (2009) 093008}.
\bibitem{12} D. Wyler, {\color{blue}Phys. Rev. D 19, 3369 (1979)}; G. C. Branco, H. P. Nilles and V. Rittenberg, {\color{blue}Phys. Rev. D 21, 3417 (1980)}.
\bibitem{13} E. Ma and G. Rajasekaran, {\color{blue}Phys. Rev. D 64, 113012 (2001) [arXiv:hep-ph/0106291]}; K. S. Babu, E. Ma and J. W. F. Valle, {\color{blue}Phys. Lett. B 552, 207 (2003) [arXiv:hep-ph/0206292]}; E. Ma, {\color{blue}Phys. Rev. D 70, 031901 (2004) [arXiv:hep-ph/0404199].}
\bibitem{14} G. Altarelli and F. Feruglio, {\color{blue}Nucl. Phys. B 741, 215 (2006) [arXiv:hep-ph/0512103].}
\bibitem{15} X. G. He, Y. Y. Keum and R. R. Volkas, {\color{blue}JHEP 0604, 039 (2006) [arXiv:hep-ph/0601001].}
\bibitem{16}  I. de Medeiros Varzielas, S. F. King and G. G. Ross, {\color{blue} Phys. Lett. B 644, 153 (2007) [arXiv:hep-ph/0512313].}
\bibitem{17}  M. Hirsch, S. Morisi and J. W. F. Valle, {\color{blue}Phys. Rev. D 78, 093007 (2008) [arXiv:0804.1521 [hep-ph]].}
\bibitem{18} KamLAND collaboration, Precision Measurement of Neutrino Oscillation Parameters with KamLAND, {\color{blue}Phys. Rev. Lett. 100 (2008) 221803 [arXiv:0801.4589] [ IN SPIRE ]}; KamLAND collaboration, Constraints on $ \theta_{13} $ from A Three-Flavor Oscillation Analysis of Reactor Antineutrinos at KamLAND, {\color{blue}Phys. Rev. D 83 (2011) 052002 [arXiv:1009.4771] [ IN SPIRE ]}; KamLAND collaboration, Reactor On-Off Antineutrino Measurement with KamLAND, {\color{blue}Phys. Rev. D 88 (2013) 033001 [arXiv:1303.4667] [ IN SPIRE ]}. 
\bibitem{19} M. Dentler, A. Hernández-Cabezudo, J. Kopp, M. Maltoni and T. Schwetz, Sterile neutrinos or flux uncertainties — Status of the reactor anti-neutrino anomaly, {\color{blue}JHEP 11 (2017) 099 [arXiv:1709.04294] [ IN SPIRE ]}; S. Gariazzo, C. Giunti, M. Laveder and Y.F. Li, Model-independent $ \bar{\nu_{e}} $ short-baseline oscillations from reactor spectral ratios, {\color{blue}Phys. Lett. B 782 (2018) 13 [arXiv:1801.06467] [ IN SPIRE ].}
\bibitem{20} P.F. de Salas, D.V. Forero, S. Gariazzo, P. Martínez-Miravé, O. Mena, C.A. Ternes, M. Tórtola, J.W.F. Valle, \textit{global reassessment of the neutrino oscillation picture}; {\color{blue}JHEP 2102 (2021) 071,  arXiv:2006.11237 [hep-ph].}
\bibitem{14nov} MEG Collaboration, Manuel Meucci, Nuovo Cim.C 43(2020) 48, arXiv:1912.08656.
\bibitem{46}  W. Grimus, Theory of Neutrino Masses and Mixing, {\color{blue}Phys.Part.Nucl.,42, pp. 566 576, 2011}.
\bibitem{newc} S.A. Abel, Subir Sarkar, P.L.White, On the cosmological domain wall problem for the minimally extended supersymmetric standard model, {\color{blue}Nucl.Phys.B 454 (1995), hep$ - $ph$ / $9506359[hep-ph].}
\bibitem{50}  F. Feruglio, C. Hagedorn, and R. Ziegler, Lepton Mixing Parameters from Discrete and CP Symmetries, {\color{blue}JHEP , 1307,  027, 2013}.
\bibitem{ra}  M.$-$C. Chen, J. Huang, K. Mahanthappa, and A. M. Wijangco, {\color{blue}JHEP 1310,  112, 2013}.
\bibitem{pat} Y. BenTov, X.-G. He, and A. Zee, "An $ A_{4}\times Z_{4} $ model for neutrino mixing," {\color{blue}JHEP 1212,  093, 2012.}
\bibitem{pc} M.C. Gonzalez-Garcia, M. Yokoyama, \textit{neutrino Masses, Mixings and oscillations}, {\color{blue}in review of PDG-2019.}
\bibitem{GGG} Kalpana Bora, Gayatri Ghosh, Debajyoti Dutta, Octant Degeneracy and Quadrant of Leptonic CPV Phase at Long Baseline $ \nu $ Experiments and Baryogenesis, {\color{blue}Adv.High Energy Phys. 2016 (2016) 9496758.}
\bibitem{GG1} Gayatri Ghosh, Kalpana Bora, Resolving Entanglement of CPV Phase with Octant of $ \theta_{23} $, and Leptogenesis, {\color{blue}Springer Proc.Phys. 174 (2016) 287-29.}
\bibitem{GG2} Naoyuki Haba, Yukihiro Mimura, Toshifumi Yamada , $ \theta_{23} $ octant measurement in 3+1 neutrino oscillations in T2HKK, {\color{blue}Phys.Rev. D101 (2020) 7, 075034.}
\bibitem{GG3} T2K Collaboration (K. Abe et al.), Measurement of neutrino and antineutrino oscillations by the T2K experiment including a new additional sample of $ \nu_{e} $interactions at the far detector {\color{blue} Phys.Rev. D96 (2017) no.9, 092006}, Erratum:{\color{blue} Phys.Rev. D98 (2018) no.1, 019902, arXiv:1707.01048 [hep-ex].}



\end{thebibliography}
\end{document}